\documentclass[12pt]{article}
\usepackage{amssymb}
\usepackage{amsthm}
\usepackage{epic,eepic}
\theoremstyle{plain}
\newtheorem{theorem}{Theorem}[section]
\newtheorem{definition}[theorem]{Definition}

\topmargin=-\headheight
\font\1=cmss10
\def\C{\mathcal{C}}
\def\Aut{\mathrm{Aut}}

\def\ff#1#2{$\scriptscriptstyle{#1\atop#2}$}

\begin{document}

{\noindent \it International Journal of Theoretical Physics\/\rm, 
\bf 39\rm, 2381--2406 (2000).}

\vbox to 2.6truecm{\vfill}

\baselineskip=13pt
\begin{flushleft}
{\Large {\bf Isomorph-Free Exhaustive Generation of
\\[2mm]
Greechie Diagrams and Automated Checking of
\\[4mm]
Their Passage by Orthomodular Lattice Equations}}\\[14.8truemm]

{\large Brendan D.~McKay$^\dag$$\footnote{
E-mail: bdm@cs.anu.edu.au; Web page:
 http://cs.anu.edu.au/$\>\tilde{}$ bdm}$,
Norman D.~Megill$^\ddag$$\footnote
{E-mail: nm@alum.mit.edu;
Web page: http://www.shore.net/$\>\tilde{}$
ndm/java/mm.html}$, and
Mladen Pavi\v ci\'c$^*$$\footnote{E-mail: mpavicic@faust.irb.hr;
Web page: http://m3k.grad.hr/pavicic}$}\\[7truemm]
$^\dag$Dept.~of Computer Science, Australian National Univ.,
Canberra, ACT, 0200, Australia.\\
$^\ddag$Boston Information Group, 30 Church St.,
Belmont MA 02478, U.~S.~A.\\
$^{*}$Dept.~of Physics, Univ.~of Maryland Baltimore County, Baltimore,
MD 21250, U.~S.~A.\\
and University of Zagreb, Gradjevinski Fakultet, Ka\v ci\'ceva 26, HR-10000 
Zagreb, Croatia. 
\end{flushleft}
\vskip25truemm
\baselineskip=12pt
\parindent=0pt

{\small\it Abstract.\/ \rm
We give a new algorithm for generating Greechie diagrams with
arbitrary chosen number of atoms or blocks (with $2,3,4,\ldots\,$ atoms)
and provide a computer program for generating the diagrams.
The results show that the previous algorithm does not produce every
diagram and that it is at least $10^5$ times slower. We also provide
an algorithm and programs for checking of Greechie diagram passage by
equations defining varieties of orthomodular lattices and give
examples from Hilbert lattices. At the end we discuss some
additional characteristics of Greechie diagrams.}
\\[5mm]
{\small\bf Keywords: \rm orthomodular lattices, Greechie diagrams,
exhaustive combinatorial generation,\break Hilbert space equations}

\vskip13truemm

\parindent=20pt\hangindent=0pt
\normalsize

\vfill\eject

\baselineskip=15pt

\section{Introduction}
\label{sec:intro}

To arrive at a Hilbert space representation of measurements
starting from plausible ``physical'' axioms has been a dream of
many physicists and mathematicians for almost seventy years.
Of course, one could not expect to recognize the axioms from
nothing but experimental data because they---provided they
exist---must be rather involved. Therefore the scientists
took the opposite road by starting with Hilbert space and
trying to read off essential mathematical properties so as to
be able to eventually simplify them and arrive at the simple
physically plausible axioms.

The first breakthrough along the \it opposite-road\/ \rm was made
by Birkhoff and von Neumann in 1936 \cite{birk-v-neum} who
recognized that a modular lattice which can be given a physical
background underlies every finite dimensional Hilbert space. In the
early sixties Mackey \cite{mackey} (and Zierler \cite{zierler})
arrived at six axioms for a poset (partially ordered set) of
physical observables which he essentially \it read off\/ \rm
from the Hilbert space properties.
In an additional famous \it seventh\/ \rm axiom, he then \it
postulated\/ \rm that the latter poset be isomorphic to the one
of the subspaces of an infinite-dimensional Hilbert space.
A few years later Piron\cite{piron}, MacLaren \cite{maclaren},
Amemyia and Araki \cite{araki}, and M{\c a}czy{\'n}ski \cite{maczin}
starting with such a poset with infima and suprema of every
two-element subset (lattice) and using similar axioms, proved
that the lattice---usually called the {\em Hilbert lattice}---is
isomorphic to a pre-Hilbert space. That enabled M{\c a}czy{\'n}ski
\cite{maczin} to postulate only the kind of a field over which the
Hilbert space should be formulated: he chose the complex one.
The Hilbert lattice is then a lattice of the subspaces of the
Hilbert space.

At that time it seemed that only two other fields could have been
postulated: the real and the quaternionic ones. But in the
early eighties Keller \cite{keller} showed that there are
other non-standard (non-archimedean) fields over which a
Hilbert space can be defined. Also, the axioms themselves proved
to be too complicated to be given plausible physical support or
simplified. Thus, the whole project lost its appeal and the majority
of researchers left the field. However, in 1995 Maria Pia Sol\`er
\cite{soler} proved that an infinite dimensional Hilbert space can
only be defined over either real, or complex, or quaternionic field
(i.e., that only finite dimensional ones allow non-standard
fields).

The latter result renewed the interest in the problem of
reconstructing the Hilbert space from an algebra of observables.
\cite{holl95,prestel,mayet98,dvurecen96,dvurecen98,mphpa98}
Also, recently devised quantum computers prompt for
such a reconstruction from an algebra which is in the field of
quantum computing usually called quantum logic in analogy to
classical logic underlying classical computers. In particular,
if we wanted quantum computers to function as quantum simulators,
i.e., to directly simulate quantum systems through their description
in the Hilbert space, we apparently have to start from such an algebra.
For, this would be the only presently conceivable way of typing in the
Hamiltonian at the console of the quantum simulator. This also means
that we have to go around the present standard axioms for the Hilbert
lattice not any more because they are complicated and physically
non-grounded but because they include universal and existential
quantifiers which are unmanageable by a quantum computer.
A way to do so would be to find lattice equations as substitutes
for the axioms. Hilbert lattice satisfies not only the
orthomodularity equation but a number of other equations as
well. Thus, if we started with a such a lattice---which can
easily be physically supported by e.g.~a quantum computer
design---we could obviously simplify the axioms and possibly
ultimately dispense with them. The problem is that only two
groups of the equations satisfied by Hilbert lattices---i.e.,
in any Hilbert space---have been found so far. We do not know
whether the Hilbert space equations form a recursively enumerable set,
i.e., whether we can determine them all. What we can do however is to
try to find as many such equations as possible and group them
according to their recursive algorithms. Each such equation can
simplify the present axioms.

However, since already equations with 4 variables contain at least
about 30 terms which one cannot further simplify, a proper tool
for finding and handling the equations is indispensable. As a
great help came Greechie diagrams (condensed Hasse diagrams) which
we will define precisely later on. E.g., to find that two equations
cannot be inferred from each other it suffices to find two
Greechie lattices which the equations interchangeably pass and fail.
As an illustration of how ``easily'' one  can find a
lattice without a computer program we cite Greechie himself:
``[In 1969] a student, beginning his dissertation, found such
a lattice. It was terribly complicated and had about eighty atoms.
The student left school and the example was lost. I've been looking
for one ever since. Recently [in 1977!] I found one.''
\cite{greechie78} So, we need an algorithm for finding Greechie
diagrams and another for finding whether a particular equation
passes or fails them. In this paper we give both. They would
not only support the afore-mentioned project of obtaining the
Hilbert space from physically plausible axioms but would also
serve for obtaining new equations in the theory of Hilbert spaces.

The first attempt at automated generation of Greechie diagrams
was made in the early eighties by G.~Beuttenm\"uller a former
student of G.~Kalmbach. \cite[pp.~319-328]{kalmb83} The
algorithm itself is not given in the book but G.~Beuttenm\"uller
kindly sent us the listing of its translation into Algol.
We rewrote it in C, and with a fast PC it took about 27 days to generate
Greechie diagrams with 13 blocks.  We estimated it would take around a
year for 14 blocks and half a century for 15 blocks, so we looked for
another approach.

The technique of \it isomorph-free exhaustive generation\/ \rm
\cite{mckay98} of Greechie diagrams gave us not only a
tremendous speed gain---48 seconds, 6 minutes, 51 minutes,
8 hours and 122 hours for 13--17 blocks, respectively (for a PC
running at 800 MHz)---but also essentially new results:
Beuttenm\"uller's algorithm must be at least incomplete since the
numbers of non-isomorphic Greechie diagrams in Kalmbach's book
(\cite{kalmb83}, p.~322) are wrong. In Sec.~\ref{sec:generation}
we give the algorithm for the above generation.

In Sec.~\ref{sec:latticeg} we give an algorithm for checking
whether a particular equation fails or passes in Greechie diagrams
provided by the algorithm from Sec.~\ref{sec:generation}.
The algorithm has helped us to find new equations that hold in
any infinite dimensional Hilbert space. \cite{mpoa99}

\section{Isomorph-free exhaustive generation of Greechie diagrams}
\label{sec:generation}

The following definitions and theorem we take over from Kalmbach
\cite{kalmb83} and Svozil and Tkadlec \cite{svozil-tkadlec}.
Definitions in the framework of \it quantum logics\/ \rm
($\sigma$-orthomodular posets) the reader can find in the book of
Pt\'ak and Pulmannov{\'a}. \cite{ptak-pulm}

\begin{definition}\label{D:diagram}
A\/ {\em diagram} is a pair $(V,E)$, where $V\ne\emptyset$ is a set of\/
{\em atoms} (drawn as points) and
$E\subseteq {\rm exp}\,V\>\backslash\,\{\emptyset\}$ is a set of\/
{\em blocks} (drawn as line segments connecting corresponding points).
A\/ {\em loop} of order $n\ge 2$ ($n$ being a natural number) in a
diagram $(V,E)$ is a sequence $(e_1,\dots e_b)\in E^n$ of mutually
different blocks such that there are mutually distinct
atoms $\nu_1,\dots,\nu_n$ with $\nu_i\in e_i\cap e_{i+1}\
(i=1,\dots,n,\ e_{n+1}=e_1)$.
\end{definition}

\begin{definition}\label{D:greechie-diagram}
A\/ {\em Greechie diagram} is a diagram satisfying the following conditions:
\begin{enumerate}
\item[(1)] Every atom belongs to at least one block.
\item[(2)] If there are at least two atoms then every block is at least
2-element.
\item[(3)] Every block which intersects with another block is at least
3-element.
\item[(4)] Every pair of different blocks intersects in at most one atom.
\item[(5)] There is no loop of order 3.
\end{enumerate}
\end{definition}

\begin{theorem}\label{th:loop-lemma}
For every Greechie diagram with only finite blocks there is exactly
one (up to an isomorphism) orthomodular poset such that there are
one-to-one correspondences between atoms and atoms and between
blocks and blocks which preserve incidence relations.
The poset is a lattice if and only if the Greechie diagram has
no loops of order~4.
\end{theorem}

In the literature, a block is also called an {\em edge\/} and an
atom is also called a {\em vertex\/} or {\it node}. (However,
we reserve the term {\it node} for an element of a Hasse diagram.)

{}From the above definitions it is clear that a block
can have not only 3 atoms but also 2 or 4 or more atoms.
However, practically all examples of Greechie diagrams used in
lattice theory are nothing but pasted 3-atom blocks. We are aware
of only two important contributions containing 4-atom blocks
(two proofs of the existence of finite lattices admitting no states
given  in \cite[Fig.~2.4.5, p.~37]{ptak-pulm} and
\cite[Fig.~17.3, p.~275]{kalmb83} and of only one result
for $n$ and $\infty$ giving orthomodular lattices without states
\cite[Fig.~17.4, p.~275]{kalmb83}.

For this reason, we have initially focussed on generation of diagrams
with every block having size~3.  Nevertheless, our description
of the generation algorithm will allow larger blocks in
anticipation of the next version of our generation program.
Our program for checking equations already handles large blocks.
Also, we are interested only in the diagrams
which correspond to lattices, i.e., only in those containing no loops
of order~4. Since the condition (5) of Def.~\ref{D:greechie-diagram}
states that there are no loops of order 3, this means that we are
interested only in diagrams with loops of order 5 and higher.
Those 3-atom Greechie diagrams which correspond to lattices we call
{\it Greechie-3-L diagrams}.

A diagram is {\it connected\/} if, for each pair of atoms $\nu, \nu'$,
there is a sequence of blocks $e_1,e_2,\ldots,e_k$ such that $\nu\in
e_1$, $\nu'\in e_k$ and $e_i\cap e_{i+1}\ne\emptyset$ for $1\le i\le
k{-}1$.  In Section~\ref{sec:latticeg} we will illustrate how the
properties of unconnected diagrams are not necessarily a simple
combination of the properties of their connected components, so our
algorithms will handle both connected and unconnected diagrams.  An
{\it isomorphism\/} from a diagram $(V_1,E_1)$ to a diagram $(V_2,E_2)$
is a bijection $\phi$ from $V_1$ to $V_2$ such that $\phi$ induces a
bijection from $E_1$ to $E_2$.  The isomorphisms from a diagram $D$ to
itself are its {\it automorphisms}, and together comprise its {\it
automorphism group\/} $\Aut(D)$.

If $D=(V,E)$ is a diagram and $e\in E$, then $D-e$ is the diagram
obtained from $D$ by removing $e$ and also removing any atoms
that were in $e$ but in no other block.  Conversely, if $e$ is a
set of atoms (not necessarily all of them atoms of $D$), then
$D+e$ is the diagram $(V\cup e,E\cup\{e\})$.  Clearly $(D+e)-e=D$.

We will describe the generation algorithm in some generality to
assist future applications.  Suppose that $\C$ is some class of
diagrams closed under isomorphisms (for example, connected
Greechie-3-L diagrams).  If $D=(V,E)\in\C$ and $|E|>1$, there
may be some $e\in E$ such that $D-e\in\C$.  If there is no such
block, we call $D$ {\it irreducible}.
It is obvious that all diagrams in $\C$
can be made from the irreducible diagrams in $\C$ by adding a
sequence of blocks one at a time, all the while staying in~$\C$.
Such a sequence of diagrams is a {\it construction path\/} for~$D$.

The basic idea behind our algorithm is to prune the set of construction
paths until (up to isomorphism) each diagram in $\C$ has exactly
one construction path.  This is achieved by two techniques acting
in consort.

The first technique is to avoid equivalent extensions.
Suppose $D\in\C$ and $e_1, e_2$ are such that $D+e_1,D+e_2\in\C$.
$D+e_1$ and $D+e_2$ are called {\it equivalent extensions of $D$}
if $|e_1|=|e_2|$ and there is an
automorphism of $D$ which maps $V\cap e_1$ onto $V\cap e_2$.
It is easy to see that the equivalence of $e_1$ and $e_2$
implies the isomorphism of $D+e_1$ and $D+e_2$, by an
isomorphism that takes $e_1$ onto $e_2$, so we do not
lose any isomorphism types of diagram if we make only one of them.

The second technique is somewhat more complicated.
Suppose we have a function $m$ with the following properties.
\begin{itemize}
   \item $m(\,)$ takes a single argument $D$ which is a reducible
      diagram in $\C$.  It returns a value which is an orbit of
      blocks under the action of $\Aut(D)$.
   \item $D-e\in\C$ for every $e\in m(D)$.
   \item If $D'$ is a diagram isomorphic to $D$, then there is
      an isomorphism from $D$ to $D'$ that maps $m(D)$ onto $m(D')$.
\end{itemize}
We will explain how to compute such a function $m(\,)$ later;
for now we will describe its purpose.  Take any reducible $D\in\C$
and $e\in m(D)$, then form $D-e$.  If we do the same starting
with a diagram $D'$ isomorphic to $D$---take $e'\in m(D')$,
then form $D'-e'$---the third property of $m(\,)$ implies that
$D-e$ and $D'-e'$ are isomorphic.
Thus, the function $m(\,)$ enables us to define a unique isomorphism
class, that of $D-e$ for $e\in m(D)$, as the {\it parent class\/} of
the isomorphism class of~$D$.
Since we wish to avoid making isomorphism types more
than once, we can decide to only make each diagram from its parent
class.  If we happen to make it from any other class,
we will reject it.

The result of the theory in \cite{mckay98} is that the combination of
the above two techniques results in each isomorphism class being
generated exactly once.  The precise method of combination is
given by the following algorithm.

\newpage

\begin{definition}\label{def:procedure}
{\em Isomorph-free Greechie diagram generation procedure}
\begin{enumerate}
\item[]{\bf procedure} scan $(D$ {\rm : diagram;} $\beta$ {\rm : integer}$)$
\begin{enumerate}
\item[]{\bf if} $D$ {\rm has exactly $\beta$ blocks} {\bf then}
\begin{enumerate}
{\bf output $D$}
\end{enumerate}
\item[]{\bf else}
\begin{enumerate}
\item[]{\bf for} {\rm each equivalence class of extensions} $D+e$ {\bf do}
\begin{enumerate}
\item[] {\bf if} $e\in m(D+e)$ {\bf then} scan($D+e$,$\beta$)
\end{enumerate}
\end{enumerate}
\end{enumerate}
\item[]{\bf end procedure}
\end{enumerate}
\end{definition}

\bigskip

\begin{figure}[htbp]\centering
  % Set unitlength to default in case it's not
  \setlength{\unitlength}{1pt}
  \begin{picture}(460,550)(-25,0)

   % 1#1
   \put(200,530){
      \begin{picture}(0,0)(0,0)
        \put(0,0){\line(1,0){50}}
        \put(0,0){\circle{10}}
        \put(25,0){\circle{10}}
        \put(50,0){\circle{10}}
      \end{picture}}

   % 2#1
   \put(200,370){
      \begin{picture}(0,0)(0,0)
        \put(0,0){\line(1,0){50}}
        \put(0,0){\circle{10}}
        \put(25,0){\circle{10}}
        \put(50,0){\circle{10}}
        \put(50,0){\line(0,1){50}}
        \put(50,25){\circle{10}}
        \put(50,50){\circle{10}}
      \end{picture}}

   % 3#1
   \put(100,210){
      \begin{picture}(0,0)(0,0)
        \put(0,0){\line(1,0){50}}
        \put(0,0){\circle{10}}
        \put(25,0){\circle{10}}
        \put(50,0){\circle{10}}
        \put(50,0){\line(0,1){50}}
        \put(50,25){\circle{10}}
        \put(50,50){\circle{10}}
        \put(0,0){\line(0,1){50}}
        \put(0,25){\circle{10}}
        \put(0,50){\circle{10}}
        \put(68,5){\makebox(0,0)[l]{$D_{3,1}$}}
      \end{picture}}

   % 3#2
   \put(270,210){
      \begin{picture}(0,0)(0,0)
        \put(0,0){\line(1,0){50}}
        \put(0,0){\circle{10}}
        \put(25,0){\circle{10}}
        \put(50,0){\circle{10}}
        \put(50,0){\line(0,1){50}}
        \put(50,25){\circle{10}}
        \put(50,50){\circle{10}}
        \put(50,0){\line(1,-1){35.3}}
        \put(67.7,-17.7){\circle{10}}
        \put(85.3,-35.3){\circle{10}}
        \put(68,5){\makebox(0,0)[l]{$D_{3,2}$}}
      \end{picture}}

   % 4#1
   \put(30,50){
      \begin{picture}(0,0)(0,0)
        \put(0,0){\line(1,0){50}}
        \put(0,0){\circle{10}}
        \put(25,0){\circle{10}}
        \put(50,0){\circle{10}}
        \put(50,0){\line(0,1){50}}
        \put(50,25){\circle{10}}
        \put(50,50){\circle{10}}
        \put(0,0){\line(0,1){50}}
        \put(0,25){\circle{10}}
        \put(0,50){\circle{10}}
        \put(-50,50){\line(1,0){50}}
        \put(-50,50){\circle{10}}
        \put(-25,50){\circle{10}}
      \end{picture}}

   % 4#2
   \put(115,50){
      \begin{picture}(0,0)(0,0)
        \put(0,0){\line(1,0){50}}
        \put(0,0){\circle{10}}
        \put(25,0){\circle{10}}
        \put(50,0){\circle{10}}
        \put(50,0){\line(0,1){50}}
        \put(50,25){\circle{10}}
        \put(50,50){\circle{10}}
        \put(0,0){\line(0,1){50}}
        \put(0,25){\circle{10}}
        \put(0,50){\circle{10}}
        \put(25,0){\line(0,1){50}}
        \put(25,25){\circle{10}}
        \put(25,50){\circle{10}}
      \end{picture}}

   % 4#3
   \put(200,50){
      \begin{picture}(0,0)(0,0)
        \put(0,0){\line(1,0){50}}
        \put(0,0){\circle{10}}
        \put(25,0){\circle{10}}
        \put(50,0){\circle{10}}
        \put(50,0){\line(0,1){50}}
        \put(50,25){\circle{10}}
        \put(50,50){\circle{10}}
        \put(0,0){\line(0,1){50}}
        \put(0,25){\circle{10}}
        \put(0,50){\circle{10}}
        \put(50,0){\line(1,-1){35.3}}
        \put(67.7,-17.7){\circle{10}}
        \put(85.3,-35.3){\circle{10}}
        \put(68,5){\makebox(0,0)[l]{$D_{4,3}$}}
      \end{picture}}

      % 4#4
   \put(330,50){
      \begin{picture}(0,0)(0,0)
        \put(0,0){\line(1,0){50}}
        \put(0,0){\circle{10}}
        \put(25,0){\circle{10}}
        \put(50,0){\circle{10}}
        \put(50,0){\line(0,1){50}}
        \put(50,25){\circle{10}}
        \put(50,50){\circle{10}}
        \put(50,0){\line(1,-1){35.3}}
        \put(67.7,-17.7){\circle{10}}
        \put(85.3,-35.3){\circle{10}}
        \put(50,0){\line(-1,-1){35.3}}
        \put(33.3,-17.7){\circle{10}}
        \put(14.7,-35.3){\circle{10}}
        \put(68,5){\makebox(0,0)[l]{$D_{4,4}$}}
      \end{picture}}

      % Lines joining diagrams
      \put(230,500){\line(0,-1){70}}
      \put(215,350){\line(-1,-1){65}}
      \put(240,350){\line(1,-1){65}}
      \put(115,190){\line(-1,-1){65}}
      \put(130,190){\line(0,-1){65}}
      \put(145,190){\line(1,-1){65}}
      \put(325,190){\line(1,-3){22}}

  \end{picture}
  \caption{Generation tree for connected Greechie-3-L diagrams
  \label{fig:tree}}
\end{figure}
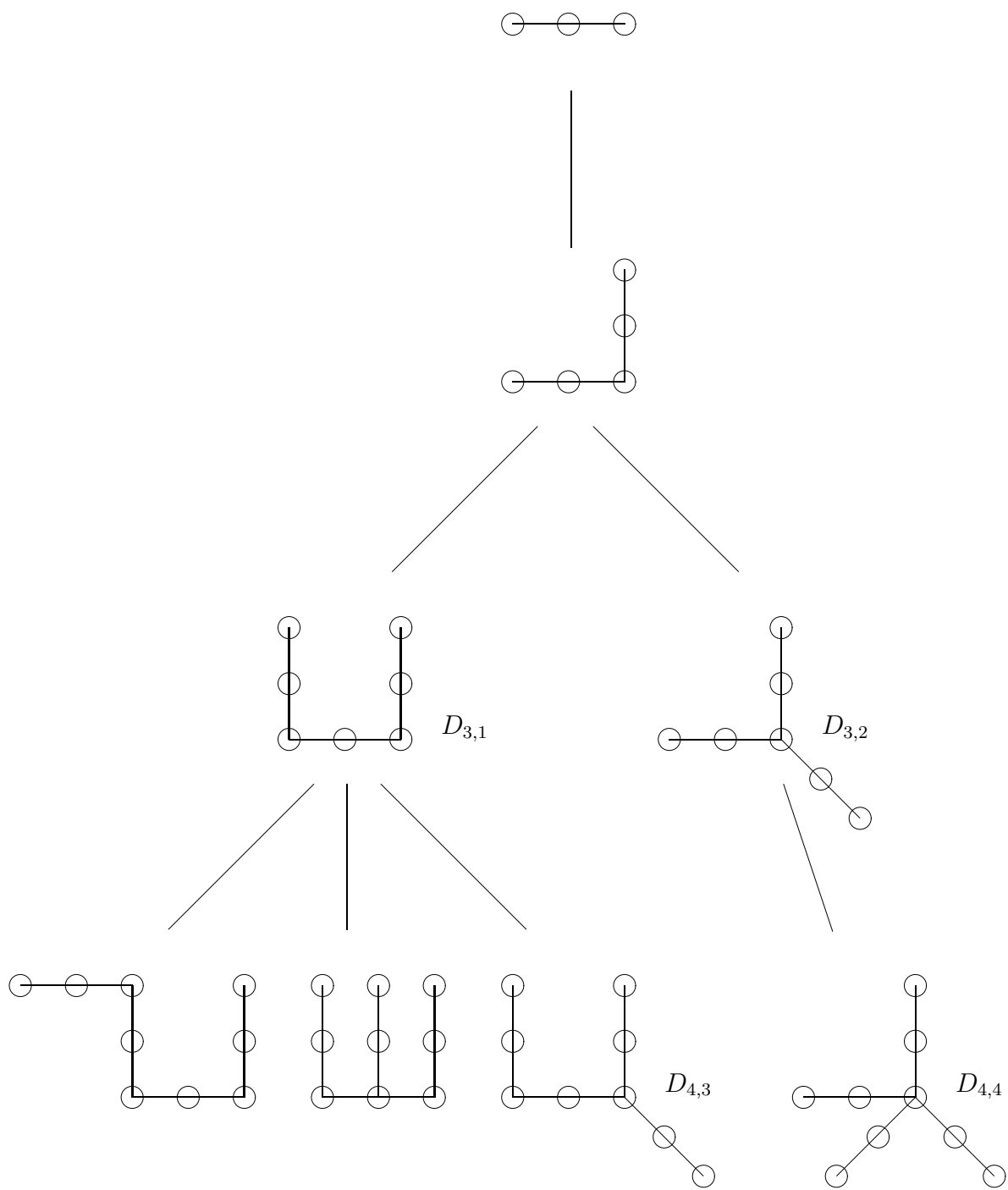

In Figure~\ref{fig:tree} we show the top four levels of the generation
tree as produced by our implementation of the algorithm for connected
Greechie-3-L diagrams.
The lines joining the diagrams show the parent-child
relationship.  Between a parent and its child, one block is added.
Note that diagram $D_{4,3}$ is made by adding a block to $D_{3,1}$,
but it could also be made by adding a block to $D_{3,2}$.
The reason that $D_{3,1}$ is its real parent is that $m(D_{4,3})$
consists of the upper right and lower right blocks (which are
equivalent) of $D_{4,3}$.  (This is a fact of our implementation which
cannot be seen by looking at the figure.)
When $D_{4,3}$ is made from $D_{3,1}$, the
new edge is seen to be in $m(D_{4,3})$ and so the diagram is accepted.
When it is made from $D_{3,2}$, the new edge is found to be not in
$m(D_{4,3})$ and so the diagram is rejected.
The idea is that each (isomorphism type of) diagram is accepted
exactly once, no matter how many times it is made.  This is proved
in the following theorem.

\begin{theorem}\label{T:procedure}
Suppose we call ${\it scan\/}(D,\beta)$ for one $D$ from each isomorphism
class of irreducible diagram in $\C$ that has at most $\beta$ blocks.
Then the output will consist of one diagram from each isomorphism
class in $\C$ with exactly $\beta$ blocks.
\end{theorem}
\begin{proof}
The theorem is a special case of one in \cite{mckay98}, and the
reader is referred to that paper for a strictly formal proof.
Here we will give a slightly less formal sketch.

Let us say that a diagram $D$ is {\it accepted\/} by the algorithm
if a call {\it scan\/}$(D,\beta)$ occurs.  We will first prove that
{\it at least\/} one member of each isomorphism class of diagram in
$\C$ with at most
$\beta$ blocks is accepted.  Then we will prove that {\it at most\/} one
member of each isomorphism class is accepted.  These two facts
together will obviously imply the truth of the theorem.

Suppose that the first assertion is false: there is an isomorphism
class in $\C$, with at most $\beta$ blocks, that is never accepted.
Let $D$ be a member of such a missing isomorphism class which has the
least number of blocks.  $D$ cannot be irreducible, since all
irreducible diagrams are accepted explicitly.  Thus, we can choose
$e\in m(D)$ and consider $D-e$.  Since $D-e\in\C$ and $D-e$ has fewer
blocks than $D$, at least one isomorph $D'$ of $D-e$ is accepted.

The isomorphism
from $D-e$ to $D'$ maps $V(D)\cap e$ onto some subset of $V(D')$.
Let $e'$ be a set of atoms consisting of that subset plus enough
new atoms to make $e'$ the same size as~$e$.  The {\bf for} loop
considers some extension $D'+e''$ equivalent to $D'+e'$, since it
considers all equivalence classes of extensions.  Moreover, since
$e\in m(D)$ we can infer that $e'\in m(D'+e')$ and consequently
that $e''\in m(D'+e'')$.  This means that the algorithm will
perform the call {\it scan\/}$(D'+e'',\beta)$, which is a contradiction
as $D'+e''$ is isomorphic to $D$ and the isomorphism class of $D$
was supposed to be not accepted at all.  This proves that all
isomorphism classes are accepted at least once.

Next suppose that some isomorphism type is accepted twice.
Namely, there two isomorphic but distinct diagrams $D$ and $D'$
in $\C$, with at most $M$ blocks, such that both $D$ and $D'$
are accepted.  Choose such a pair $D, D'$ with the least
number of blocks.

As before, $D$ and $D'$ cannot be irreducible, so they must
be accepted by some calls {\it scan\/}$((D-e)+e,\beta)$ and
{\it scan\/}$((D'-e')+e',\beta)$ which arise from the calls
{\it scan\/}$(D-e,\beta)$ and {\it scan\/}$(D'-e',\beta)$, respectively,
where $e\in m(D)$ and $e'\in m(D')$.  The properties of
$m(\,)$ ensure that $D-e$ and $D'-e'$ are isomorphic, so they
must in fact be the same diagram $D''$ (since isomorphism classes with
fewer blocks than $D$ are accepted at most once by assumption).
However, $D''+e$ and $D''+e'$ are equivalent but distinct extensions
of $D''$, which violates the {\bf for} loop specification.
This contradiction completes the proof.
\end{proof}

\medskip

The success of the algorithm requires us to
be able to find the irreducible diagrams in $\C$ by some other
method, but in many important cases this is easy.  We give the most
important example.

\begin{theorem}\label{T:irred}
 Suppose $\C$ is a class of Greechie diagrams defined by some
 fixed set of permissible block sizes, some fixed set of permissible
 loop lengths, and an optional restriction to
 connected diagrams.
 Then the only irreducible diagrams in $\C$ are those with one block.
\end{theorem}

\begin{proof}
Consider a diagram $D\in\C$ with more than one block.

If $\C$ is not restricted to connected diagrams, $D-e\in\C$ for
any $e\in E(D)$, so $D$ is reducible.

Suppose instead that $\C$ contains only connected diagrams.
Choose a longest possible sequence $S$ of distinct blocks
$e_1,e_2,\ldots,e_k$, where $e_i\cap e_{i+1}\ne\emptyset$
for $1\le i\le k{-}1$.
Let $\nu_1$ and $\nu_2$ be two atoms of $D-e_k$.  Since $D$ is connected,
there is a chain of blocks from $\nu_1$ to $\nu_2$.  This same chain
is in $D-e_m$ unless it contains $e_k$.  However, all the blocks $D$
intersecting $e_k$ are in $S$ (or else $S$ can be made longer), so
$e_k$ can be replaced in $S$ by some portion of~$S$.
Hence $D-e_k$ is connected, so $D$ is reducible.
\end{proof}

\medskip

The correctness of the algorithm does not depend on the definition
of $m(\,)$ provided it has the properties we required of it.  The
actual definition of $m(\,)$ used in our program is carefully tuned
for optimal observed performance, and is too complicated to describe
here in detail, but we will outline a simpler definition that
is the same in essence.

The key to our implementation of $m(\,)$ is the first author's
graph isomorphism program {\tt nauty} \cite{mckay90} can be used.
{\tt nauty} takes
a simple graph $G$, perhaps with colored atoms, and produces two
outputs.  One is the automorphism group $\Aut(G)$, in the form of a
set of generators.  The other is a canonical labelling of $G$,
which is a graph $c(G)$ isomorphic to $G$.  The function $c$ is
``canonical'' in the sense that $c(G)=c(G')$ for every graph $G'$
isomorphic to~$G$.  To apply {\tt nauty} to a diagram $(V,E)$,
we can use the incidence graph $G=(V\cup E,\{(v,e) | v\in e\})$.

The generators
for $\Aut(G)$ can be easily converted into generators for $\Aut(D)$
and then used to determine the equivalence classes of extensions.
This enables us to implement the requirement of avoiding
equivalent extensions.

The canonical labelling $c(\,)$ produced by {\tt nauty} enables us
to define $m(\,)$.
Take the block $e$ such that $D-e\in\C$ and $e$ is given the
least new label by $c(\,)$.
If we define $m(D)$ to be the orbit of blocks that contains~$e$,
we find that the three requirements we imposed on $m(\,)$ are
satisfied.

As we have said, our real program uses a more complex definition of
$m(\,)$.  We do not use the incidence graph $G$, but instead use a
prototype variant of {\tt nauty} that operates on diagrams directly.
Since our program makes connected diagrams, we took $m(D)$ to be an
orbit of feet if there were any, where a foot is a block with only
one atom that also lies in other blocks.
This avoids many connectivity tests, since removal of
a foot necessarily preserves connectivity.
It also avoids many futile
extensions: adding a non-foot $e$ must be done in such a way that
any existing feet become non-feet, as otherwise $e\notin m(D+e)$.

In order to generate only the connected Greechie-3-L diagrams having
$M$ blocks but no feet, a reasonable approach is to generate all the
diagrams, having feet or not, with $M-1$ blocks first.  Then the
$M$-th block can be added in such a way that uses at least 2 of the
existing atoms and also turns any feet into non-feet.  It is
also possible to make a generator that makes foot-free diagrams
while staying entirely within that class, but it does not appear
likely to be much different in efficiency.

\medskip
\noindent
{\bf Program} {\tt greechie}. Our implementation of the algorithm
is a self-contained program called {\tt greechie}\footnote{{\tt
ftp://m3k.grad.hr/pavicic/greechie}, {\tt
http://cs.anu.edu.au/$\>\tilde{}\>$bdm/nauty/greechie.html}.
Many of the diagrams computed with the program are also available
at those places.}
that takes as parameters the number of blocks, an optional upper
bound on the number of atoms, and whether or not feet are
permitted.  It then produces one representative of each isomorphism
class of connected Greechie-3-L diagram with those properties.
The diagrams can then be processed as they are generated, with no
need to store them.  There is also an option for dividing the set
of diagrams into disjoint subsets, and efficiently producing only
one of the subsets.  This allows long computations to be broken
into manageable pieces that can be run independently, even on
different computers, without much change to the total running time.

In Table~\ref{T:tableA} we list the numbers of Greechie-3-L
diagrams for
small values of $\alpha$ and $\beta$.  In each cell of the tables,
the upper value is the total
number of connected Greechie-3-L diagrams, and the lower value is the
number of those which have no feet.  Both counts are 0 if the table
cell is empty.  The table includes all
possible values of $\beta$ for $\alpha\le 29$ and all possible values
of $\alpha$ for $\beta\le 17$.

\begin{table}[p]
{\small\begin{center}
%
%-v=3-20 e=1-14--------
\begin{tabular}{|c|cccccccccccccc|c|}
\hline
$\alpha\,\setminus\,\beta$&1&2&3&4&5&6&7&8&9&10&11&12&13&14&\it total\\
\hline
3&\ff{1}{1}&&&&&&&&&&&&&&\ff
  {1}{1}\\
5&&\ff{0}{1}&&&&&&&&&&&&&\ff
  {0}{1}\\
7&&&\ff
   {0}{2}&&&&&&&&&&&&\ff
  {0}{2}\\
9&&&&\ff{0}{4}&&&&&&&&&&&\ff
  {0}{4}\\
10&&&&&\ff{1}{1}&&&&&&&&&&\ff
  {1}{1}\\
11&&&&&\ff{0}{8}&&&&&&&&&&\ff
  {0}{8}\\
12&&&&&&\ff
   {1}{3}&&&&&&&&&\ff
  {1}{3}\\
13&&&&&&\ff
   {0}{19}&\ff{2}{2}&&&&&&&&\ff
  {2}{21}\\
14&&&&&&&\ff{1}{14}&\ff{1}{1}&&&&&&&\ff
  {2}{15}\\
15&&&&&&&\ff{0}{48}&\ff{6}{16}&\ff
   {1}{1}&\ff{1}{1}&&&&&\ff
  {8}{66}\\
16&&&&&&&&\ff{1}{62}&\ff
   {12}{15}&\ff{1}{1}&&&&&\ff
  {14}{78}\\
17&&&&&&&&\ff{0}{126}&\ff
   {11}{119}&\ff{21}{24}&\ff{2}{3}&&&&\ff
  {34}{272}\\
18&&&&&&&&&\ff
   {1}{281}&\ff{71}{209}&\ff{27}{31}&\ff
   {4}{4}&&&\ff
  {103}{525}\\
19&&&&&&&&&\ff
   {0}{355}&\ff{19}{819}&\ff{261}{490}&\ff
   {67}{84}&\ff{6}{6}&&\ff
  {353}{1754}\\
20&&&&&&&&&&\ff{1}{1239}&\ff{251}{2347}&\ff
   {834}{1217}&\ff{147}{166}&\ff{1}{1}&\ff
  {1234}{4970}\\
\hline
\it total&\ff{1}{1}&\ff{0}{1}&\ff
   {0}{2}&\ff{0}{4}&\ff{1}{9}&\ff
   {1}{22}&\ff{3}{64}&\ff{8}{205}&\ff
   {25}{771}&\ff{114}{3330}&\ff{571}{16571}&\ff
   {3675}{95327}&\ff{27687}{628555}&\ff{239844}{4713887}&\\
\hline
\end{tabular}

\smallskip

%-v=21-35 e=10-17--------
\begin{tabular}{|c|cccccccc|c|}
\hline
$\alpha\,\setminus\,\beta$&10&11&12&13&14&15&16&17&\it total\\
\hline
21&\ff{0}{1037}&\ff{29}{5199}&\ff
   {2052}{8273}&\ff{2884}{3872}&\ff{405}{437}&\ff
   {4}{4}&&&\ff
  {5374}{18822}\\
22&&\ff{1}{5377}&\ff
   {675}{22219}&\ff{12849}{31805}&\ff{10885}{13440}&\ff
   {905}{908}&\ff{4}{4}&&\ff
  {25319}{73753}\\
23&&\ff{0}{3124}&\ff
   {42}{30923}&\ff{10235}{109467}&\ff{74698}{141454}&\ff
   {45905}{53075}&\ff{2837}{2861}&\ff{17}{17}&\ff
  {133734}{340921}\\
24&&&\ff
   {1}{22931}&\ff{1508}{182581}&\ff{113376}{588877}&\ff
   {435636}{695281}&\ff{207767}{225862}&\ff{10723}{10756}&\ff
  {769050}{1726327}\\
25&&&\ff
   {0}{9676}&\ff{57}{173841}&\ff{37406}{1194496}&\ff
   {1061800}{3508197}&\ff{2670655}{3787093}&\ff{1030296}{1091395}&\ff
  {4849700}{9814445}\\
26&&&&\ff{1}{96213}&\ff{2997}{1346088}&\ff
   {664386}{8429567}&\ff{9287661}{22797227}&\ff{17387077}{22548822}&\ff
  {33292775}{61419172}\\
27&&&&\ff{0}{30604}&\ff{75}{932262}&\ff
   {110376}{11227170}&\ff{9494774}{64431928}&\ff{80621116}{161294913}&\ff
  {248439348}{423756705}\\
28&&&&&\ff{1}{398359}&\ff
   {5463}{9117775}&\ff{2905751}{99744819}&\ff{120414885}{530135685}&\ff
  {2009490610}{3203436511}\\
29&&&&&\ff{0}{98473}&\ff
   {95}{4807157}&\ff{280150}{94035080}&\ff{60035427}{945730356}&\ff
  {17618049369}{26495317590}\\
30&&&&&&\ff
   {1}{1630602}&\ff{9329}{57743948}&\ff{10280997}{1021031346}&?\\
31&&&&&&\ff
   {0}{321572}&\ff{118}{23993678}&\ff{636294}{719130759}&?\\
32&&&&&&&\ff{1}{6612148}&\ff{15096}{346356783}&?\\
33&&&&&&&\ff{0}{1063146}&\ff{143}{116542189}&?\\
34&&&&&&&&\ff{1}{26603735}&?\\
35&&&&&&&&\ff{0}{3552563}&?\\
\hline
\it total&\ff{114}{3330}&\ff{571}{16571}&\ff
   {3675}{95327}&\ff{27687}{628555}&\ff{239844}{4713887}&\ff
   {2324571}{39791308}&\ff{24859047}{374437794}&\ff{290432072}{3894029319}&\\
\hline
\end{tabular}

\smallskip

%-v=24-29 e=18-24--------
\begin{tabular}{|c|ccccccc|c|}
\hline
$\alpha\,\setminus\,\beta$&18&19&20&21&22&23&24&\it total\\
\hline
24&\ff{39}{39}&&&&&&&\ff
  {769050}{1726327}\\
25&\ff{49350}{49611}&\ff{134}{134}&\ff
   {2}{2}&&&&&\ff
  {4849700}{9814445}\\
26&\ff{5702603}{5952608}&\ff{247469}{248066}&\ff
   {581}{581}&&&&&\ff
  {33292775}{61419172}\\
27&\ff{121218048}{147629428}&\ff{35519085}{36732179}&\ff
   {1471694}{1474036}&\ff{4180}{4185}&&&&\ff
  {248439348}{423756705}\\
28&\ff{717349032}{1237468066}&\ff{911521449}{1062867580}&\ff
   {247041933}{253438446}&\ff{10216096}{10229781}&\ff{35992}{35992}&\ff
   {8}{8}&&\ff
  {2009490610}{3203436511}\\
29&\ff{1448834695}{4690081789}&\ff{6661929716}{10298834720}&\ff
   {7447274324}{8422315646}&\ff{1916771296}{1956383755}&\ff{82563871}{82670819}&\ff
   {359550}{359550}&\ff{245}{245}&\ff
  {17618049369}{26495317590}\\
\hline
\end{tabular}
\end{center}}
\caption{Counts of connected Greechie-3-L diagrams}
\label{T:tableA}
\end{table}

\medskip

For reasons explained later, we have particular interest in those
diagrams containing close to the maximum number of blocks for a given
number of atoms.  This prompted us to compute additional near-maximal
diagrams past the size where finding all the diagrams is practical.

To keep the discussion simple, we restrict ourselves to Greechie-3-L
diagrams, not necessarily connected.  By the {\it type\/} of a diagram
we mean the pair $(\alpha,\beta)$, where $\alpha$ is the number of atoms
and $\beta$ is the number of blocks.
The {\it rank\/} of an atom is the number of blocks which contain it.

The first observation is that a diagram $D$ of type $(\alpha,\beta)$
has an atom whose rank is at most $\lfloor 3\beta/\alpha\rfloor$.
Over most of our computational
range, $\alpha<\beta$, so this value is at most~2.
If there is an atom of rank~1, we can make $D$ either by adding a foot to
a diagram of type $(\alpha{-}2,\beta{-}1)$, or by adding one block and
one atom to a diagram of type $(\alpha{-}1,\beta{-}1)$.  On the
other hand, if there is an atom of rank~2 but none of rank~1, we can make
$D$ by adding one atom and two blocks to a diagram of type
$(\alpha{-}1,\beta{-}2)$.  So, if we have already made the diagrams of types
$(\alpha{-}2,\beta{-}1)$, $(\alpha{-}1,\beta{-}1)$,
and $(\alpha{-}1,\beta{-}2)$, we can easily
extend them to make those of type $(\alpha,\beta)$, provided $\alpha<\beta$.

Sometimes the class $(\alpha{-}1,\beta{-}2)$ may be too onerous to compute.
In this case, there are two other approaches we might be able to take
to making those diagrams of type $(\alpha,\beta)$ whose minimum atom rank is~2.
Define $\alpha_i$ to be the number of atoms of rank $i$ (and
recall that we are assuming $\alpha_1=0$).
Counting the pairs (block, atom in block) in two ways, we have
$2\alpha_2 + 3\alpha_3 + 4(\alpha-\alpha_2-\alpha_3) \le 3\beta$.
Since also $\alpha_2+\alpha_3\le \alpha$, we have
\begin{equation}
  2\alpha_2 + \alpha_3 \ge 4\alpha-3\beta \hbox{ and }
        \alpha_2\ge 3\alpha-3\beta.\label{E:a}
\end{equation}

Now consider the case of a Greechie-3-L diagram with
$\alpha_1=0$ and $6\alpha>7\beta$.
Applying the second part of (\ref{E:a}) we find that $2\alpha_2>\beta$, which
implies that some block contains at least two atoms of rank~2.  Therefore,
we can make the diagram by adding two atoms and three blocks to a
diagram of type $(\alpha{-}2,\beta{-}3)$.

If $6\alpha\le 7\beta$ but $\alpha>\beta$,
there might be no atoms of rank~1, nor two atoms
of rank~2 in the same block.  In this case, we know that there are exactly
$2\alpha_2$ blocks containing an atom of rank~2.
The total rank of the atoms of rank greater than~3 is
$3\beta-2\alpha_2-3\alpha_3$, so we have at least
$4\alpha_2-(3\beta-2\alpha_2-3\alpha_3) = 6\alpha_2+3\alpha_3-3\beta$
pairs $(x,y)$ such
that $x$ and $y$ are atoms of rank~2 and~3, respectively, lying in the
same block.  Now, if we suppose that $14\alpha>15\beta$, we find that
$6\alpha_2+3\alpha_3-3\beta > \alpha_3$,
implying that two of the pairs $(x,y)$ have the
same~$y$.  That is, there is an atom of rank~3 lying in two blocks which
each contain an atom of rank~2.  Therefore, we can make this diagram by
adding 3 atoms and 5 blocks to a diagram of type $(\alpha-3,\beta-5)$.

\begin{table}[hbt]
{\small\begin{center}

%--near-extremal cases.  this table made by hand

\def\al{\hbox to 0pt{\hss$\scriptstyle\ge$}}
\def\ss#1{$\scriptstyle#1$}

\begin{tabular}{|c|cccccccccccc|}
\hline
$\alpha\,\setminus\,\beta$&25&26&27&28&29&30&31&32&33&34&35&36\\
\hline
30&\ss{3982}&\ss{4}&&&&&&&&&&\\
31&?&\ss{81068}&\ss{71}&\ss{1}&&&&&&&&\\
32&?&?&\al\ss{313813}&\ss{1643}&&&&&&&&\\
33&?&?&?&?&\ss{\al51643}&\ss{66}&&&&&&\\
34&?&?&?&?&?&\al\ss{185733}&\ss{2113}&\ss{19}&&&&\\
35&?&?&?&?&?&?&?&\al\ss{70035}&\ss{325}&\ss{17}&\ss{5}&\\
36&?&?&?&?&?&?&?&?&?&\al\ss{7871}&\ss{136}&\ss{1}\\
37&?&?&?&?&?&?&?&?&?&?&?&\al\ss{1693}\\
\hline
\end{tabular}

\end{center}}

\caption{Counts of large extremal Greechie-3-L diagrams}
\label{T:tableB}
\end{table}

Using these ideas, we were able to compute (see Table~\ref{T:tableB}) 
the  Greechie-3-L diagrams with up to~36 atoms  having the maximum and, 
in some cases, near-maximum possible numbers of blocks. 
%The counts of those extremal diagrams are shown in Table~\ref{T:tableB}.

The maximum number of blocks
shown in each row of the table is the maximum possible.  All the
diagrams counted in the table turned out to be connected, even though
our programs did not assume connectivity.

%-------36x36 figure------------------------------------------------

\begin{figure}[hbt]\centering
\setlength{\unitlength}{0.00054167in}
\begingroup\makeatletter\ifx\SetFigFont\undefined%
\gdef\SetFigFont#1#2#3#4#5{%
  \reset@font\fontsize{#1}{#2pt}%
  \fontfamily{#3}\fontseries{#4}\fontshape{#5}%
  \selectfont}%
\fi\endgroup%
{\renewcommand{\dashlinestretch}{30}
\let\whiten=\relax  
\begin{picture}(9208,9279)(0,-10)
\path(95,4830)(1843,4766)
\path(1003,1815)(2383,2889)
\path(6815,465)(5997,2010)
\path(8978,2770)(7348,3431)
\path(7220,8460)(6288,6980)
\path(4155,9167)(4394,7448)
\path(3680,82)(4047,1792)
\path(1347,7735)(2670,6577)
\path(9125,5932)(7460,5395)
\path(7455,5392)(7455,5393)(7454,5396)
        (7452,5401)(7449,5408)(7445,5420)
        (7440,5435)(7433,5454)(7424,5477)
        (7414,5506)(7402,5539)(7388,5577)
        (7372,5619)(7354,5667)(7334,5718)
        (7313,5773)(7289,5832)(7265,5894)
        (7239,5959)(7211,6025)(7182,6093)
        (7153,6162)(7122,6232)(7091,6301)
        (7059,6369)(7026,6436)(6993,6501)
        (6960,6563)(6926,6623)(6892,6680)
        (6858,6734)(6824,6783)(6790,6829)
        (6755,6871)(6721,6908)(6685,6940)
        (6650,6968)(6614,6990)(6577,7008)
        (6540,7020)(6502,7028)(6463,7030)
        (6422,7027)(6381,7019)(6338,7005)
        (6293,6987)(6262,6972)(6230,6954)
        (6197,6935)(6162,6912)(6127,6887)
        (6090,6859)(6052,6829)(6012,6795)
        (5971,6759)(5929,6720)(5885,6678)
        (5839,6633)(5791,6585)(5742,6535)
        (5692,6481)(5639,6424)(5585,6365)
        (5529,6303)(5472,6238)(5413,6171)
        (5353,6101)(5291,6028)(5227,5953)
        (5162,5876)(5097,5797)(5030,5716)
        (4962,5634)(4893,5550)(4823,5464)
        (4753,5377)(4682,5290)(4612,5202)
        (4541,5114)(4470,5025)(4400,4937)
        (4331,4849)(4262,4762)(4194,4676)
        (4128,4592)(4062,4509)(3999,4428)
        (3938,4350)(3878,4274)(3821,4201)
        (3766,4131)(3714,4064)(3665,4000)
        (3619,3941)(3575,3884)(3535,3832)
        (3497,3784)(3463,3740)(3432,3699)
        (3404,3663)(3379,3631)(3357,3602)
        (3337,3577)(3321,3556)(3307,3538)
        (3296,3524)(3287,3512)(3280,3503)
        (3275,3496)(3272,3492)(3270,3489)
        (3268,3488)(3268,3487)
\path(6725,7665)(6725,7664)(6724,7662)
        (6723,7658)(6722,7651)(6720,7641)
        (6717,7628)(6713,7611)(6708,7590)
        (6702,7565)(6696,7535)(6688,7501)
        (6679,7462)(6670,7418)(6659,7370)
        (6648,7318)(6635,7261)(6622,7201)
        (6609,7136)(6595,7069)(6580,6999)
        (6565,6926)(6550,6852)(6535,6776)
        (6520,6698)(6505,6620)(6491,6542)
        (6477,6463)(6463,6385)(6450,6308)
        (6438,6231)(6427,6156)(6416,6082)
        (6407,6010)(6398,5940)(6391,5872)
        (6385,5805)(6380,5741)(6376,5678)
        (6374,5618)(6373,5559)(6374,5503)
        (6376,5448)(6379,5396)(6384,5344)
        (6391,5295)(6399,5246)(6409,5199)
        (6420,5153)(6433,5107)(6448,5062)
        (6464,5017)(6482,4972)(6502,4927)
        (6524,4882)(6548,4837)(6574,4790)
        (6603,4744)(6634,4696)(6667,4648)
        (6702,4599)(6740,4549)(6780,4497)
        (6822,4445)(6866,4392)(6912,4338)
        (6960,4282)(7009,4226)(7061,4170)
        (7113,4113)(7167,4055)(7221,3997)
        (7277,3939)(7332,3881)(7388,3824)
        (7443,3768)(7498,3712)(7552,3658)
        (7605,3605)(7656,3554)(7705,3506)
        (7752,3460)(7796,3416)(7838,3375)
        (7876,3338)(7912,3303)(7944,3272)
        (7973,3244)(7999,3219)(8021,3198)
        (8039,3180)(8055,3165)(8068,3153)
        (8077,3144)(8084,3137)(8089,3132)
        (8093,3129)(8094,3128)(8095,3127)
\path(4158,9177)(4158,9176)(4159,9175)
        (4161,9172)(4164,9168)(4168,9162)
        (4173,9154)(4180,9143)(4189,9130)
        (4199,9113)(4212,9094)(4227,9071)
        (4244,9045)(4263,9015)(4285,8981)
        (4309,8943)(4336,8902)(4365,8857)
        (4396,8808)(4430,8756)(4466,8699)
        (4504,8640)(4544,8577)(4586,8510)
        (4630,8441)(4676,8368)(4724,8293)
        (4773,8216)(4823,8136)(4874,8055)
        (4926,7971)(4979,7887)(5033,7800)
        (5087,7713)(5141,7626)(5195,7537)
        (5249,7449)(5303,7360)(5357,7271)
        (5410,7183)(5463,7095)(5515,7007)
        (5566,6921)(5616,6835)(5665,6750)
        (5713,6667)(5759,6584)(5805,6503)
        (5849,6423)(5892,6344)(5933,6266)
        (5974,6190)(6012,6115)(6050,6042)
        (6086,5969)(6120,5898)(6153,5828)
        (6185,5759)(6215,5691)(6244,5624)
        (6272,5558)(6298,5492)(6323,5427)
        (6347,5363)(6370,5299)(6392,5235)
        (6413,5171)(6433,5107)(6452,5043)
        (6471,4978)(6488,4913)(6505,4847)
        (6521,4780)(6536,4712)(6551,4643)
        (6565,4573)(6578,4501)(6591,4428)
        (6603,4353)(6615,4276)(6626,4198)
        (6637,4118)(6647,4036)(6657,3952)
        (6666,3867)(6675,3779)(6683,3690)
        (6691,3599)(6699,3506)(6706,3411)
        (6713,3315)(6720,3218)(6726,3119)
        (6732,3018)(6738,2917)(6743,2815)
        (6749,2712)(6754,2608)(6758,2504)
        (6763,2401)(6767,2297)(6771,2194)
        (6775,2092)(6779,1990)(6782,1890)
        (6785,1792)(6789,1696)(6791,1602)
        (6794,1510)(6797,1422)(6799,1336)
        (6801,1254)(6803,1175)(6805,1100)
        (6807,1030)(6808,963)(6809,901)
        (6811,843)(6812,789)(6813,740)
        (6814,695)(6815,655)(6815,620)
        (6816,588)(6816,561)(6817,538)
        (6817,518)(6817,502)(6818,489)
        (6818,480)(6818,472)(6818,467)
        (6818,464)(6818,463)(6818,462)
\path(7350,3432)(7350,3433)(7351,3436)
        (7353,3440)(7356,3448)(7360,3459)
        (7365,3474)(7372,3493)(7380,3517)
        (7390,3545)(7401,3578)(7414,3616)
        (7429,3658)(7445,3706)(7462,3757)
        (7480,3813)(7499,3873)(7519,3936)
        (7539,4001)(7560,4070)(7581,4140)
        (7601,4211)(7621,4283)(7640,4356)
        (7658,4428)(7675,4500)(7691,4570)
        (7704,4639)(7716,4706)(7726,4771)
        (7733,4834)(7738,4894)(7740,4950)
        (7740,5004)(7736,5055)(7729,5102)
        (7719,5146)(7705,5186)(7688,5223)
        (7667,5256)(7642,5286)(7613,5313)
        (7580,5337)(7543,5357)(7501,5375)
        (7455,5390)(7420,5399)(7383,5406)
        (7344,5413)(7302,5418)(7257,5422)
        (7209,5424)(7158,5425)(7104,5425)
        (7047,5424)(6987,5421)(6924,5417)
        (6857,5412)(6787,5406)(6713,5398)
        (6637,5389)(6557,5379)(6474,5368)
        (6387,5356)(6298,5343)(6205,5329)
        (6110,5313)(6012,5297)(5911,5280)
        (5808,5262)(5702,5244)(5595,5224)
        (5486,5204)(5375,5184)(5263,5163)
        (5150,5141)(5036,5119)(4923,5097)
        (4809,5075)(4696,5052)(4583,5030)
        (4472,5008)(4363,4986)(4255,4964)
        (4150,4942)(4048,4921)(3949,4901)
        (3853,4881)(3761,4862)(3673,4844)
        (3590,4826)(3511,4810)(3437,4794)
        (3368,4780)(3304,4766)(3246,4754)
        (3192,4742)(3144,4732)(3101,4723)
        (3063,4715)(3030,4708)(3002,4702)
        (2978,4697)(2959,4693)(2943,4689)
        (2931,4687)(2923,4685)(2917,4683)
        (2913,4683)(2911,4682)(2910,4682)
\path(5995,2005)(5996,2006)(5998,2007)
        (6003,2009)(6010,2014)(6020,2019)
        (6034,2027)(6052,2038)(6073,2050)
        (6099,2065)(6130,2083)(6164,2104)
        (6203,2127)(6246,2153)(6293,2181)
        (6344,2212)(6397,2245)(6454,2280)
        (6512,2318)(6573,2356)(6634,2397)
        (6697,2438)(6759,2480)(6821,2523)
        (6882,2567)(6942,2611)(7000,2654)
        (7055,2698)(7108,2742)(7158,2785)
        (7204,2828)(7247,2871)(7286,2913)
        (7320,2954)(7351,2995)(7376,3035)
        (7397,3075)(7413,3115)(7423,3154)
        (7429,3193)(7430,3232)(7425,3272)
        (7415,3311)(7399,3351)(7379,3391)
        (7353,3432)(7333,3460)(7310,3489)
        (7285,3518)(7257,3548)(7226,3579)
        (7192,3611)(7155,3643)(7116,3677)
        (7073,3711)(7027,3746)(6978,3783)
        (6926,3820)(6871,3859)(6813,3899)
        (6751,3940)(6687,3982)(6619,4025)
        (6548,4069)(6474,4115)(6398,4162)
        (6319,4209)(6237,4258)(6152,4308)
        (6065,4359)(5976,4410)(5885,4463)
        (5792,4516)(5697,4569)(5601,4623)
        (5504,4678)(5406,4732)(5307,4787)
        (5208,4842)(5109,4896)(5010,4951)
        (4912,5004)(4815,5057)(4719,5109)
        (4624,5161)(4532,5211)(4441,5259)
        (4354,5307)(4269,5352)(4187,5396)
        (4109,5438)(4034,5478)(3963,5516)
        (3896,5552)(3834,5585)(3775,5616)
        (3721,5645)(3672,5671)(3627,5695)
        (3586,5716)(3550,5735)(3518,5752)
        (3491,5767)(3467,5779)(3447,5790)
        (3431,5798)(3418,5805)(3408,5811)
        (3401,5814)(3395,5817)(3392,5819)
        (3391,5820)(3390,5820)
\path(1855,4770)(1855,4769)(1855,4766)
        (1855,4761)(1855,4753)(1855,4741)
        (1854,4725)(1854,4705)(1854,4680)
        (1854,4650)(1854,4615)(1854,4575)
        (1854,4530)(1854,4481)(1855,4426)
        (1856,4367)(1857,4305)(1859,4239)
        (1861,4170)(1863,4099)(1867,4026)
        (1870,3952)(1875,3877)(1880,3802)
        (1887,3728)(1894,3655)(1902,3583)
        (1912,3514)(1922,3447)(1934,3382)
        (1947,3321)(1962,3263)(1978,3209)
        (1996,3158)(2016,3112)(2038,3070)
        (2062,3032)(2087,2999)(2116,2970)
        (2146,2946)(2179,2926)(2215,2910)
        (2254,2899)(2296,2893)(2341,2890)
        (2389,2892)(2425,2895)(2462,2901)
        (2501,2908)(2542,2918)(2586,2930)
        (2632,2944)(2680,2960)(2731,2978)
        (2784,2999)(2840,3022)(2898,3047)
        (2959,3074)(3023,3104)(3090,3136)
        (3159,3171)(3231,3207)(3306,3246)
        (3383,3287)(3463,3329)(3546,3374)
        (3630,3421)(3718,3469)(3807,3520)
        (3898,3572)(3992,3625)(4086,3680)
        (4183,3736)(4280,3793)(4379,3850)
        (4478,3909)(4578,3968)(4678,4028)
        (4778,4087)(4877,4147)(4975,4206)
        (5073,4265)(5169,4323)(5263,4380)
        (5355,4436)(5444,4490)(5531,4543)
        (5615,4594)(5695,4644)(5772,4691)
        (5845,4735)(5914,4778)(5978,4817)
        (6038,4855)(6094,4889)(6145,4921)
        (6192,4949)(6234,4975)(6271,4999)
        (6304,5019)(6333,5037)(6358,5052)
        (6378,5065)(6395,5075)(6409,5084)
        (6419,5090)(6427,5095)(6432,5098)
        (6436,5100)(6437,5102)(6438,5102)
\path(2385,2897)(2386,2896)(2388,2894)
        (2392,2889)(2398,2882)(2406,2871)
        (2418,2858)(2433,2840)(2452,2818)
        (2474,2792)(2500,2762)(2529,2727)
        (2563,2689)(2600,2647)(2640,2601)
        (2684,2552)(2730,2501)(2779,2447)
        (2830,2391)(2883,2335)(2938,2277)
        (2993,2220)(3049,2163)(3106,2108)
        (3163,2054)(3219,2002)(3275,1953)
        (3330,1906)(3384,1863)(3436,1824)
        (3488,1788)(3538,1758)(3587,1731)
        (3634,1709)(3680,1693)(3725,1681)
        (3768,1675)(3810,1674)(3850,1679)
        (3890,1690)(3929,1706)(3967,1728)
        (4005,1756)(4043,1790)(4068,1816)
        (4093,1844)(4119,1875)(4144,1909)
        (4170,1946)(4196,1986)(4222,2030)
        (4249,2076)(4276,2127)(4304,2180)
        (4332,2237)(4361,2298)(4390,2362)
        (4420,2430)(4451,2501)(4482,2576)
        (4514,2654)(4547,2735)(4580,2819)
        (4614,2907)(4648,2998)(4683,3091)
        (4719,3188)(4755,3287)(4791,3388)
        (4828,3491)(4865,3596)(4902,3703)
        (4940,3811)(4977,3921)(5015,4031)
        (5052,4141)(5090,4251)(5127,4361)
        (5163,4470)(5199,4578)(5234,4685)
        (5269,4790)(5303,4892)(5336,4992)
        (5367,5089)(5398,5182)(5427,5272)
        (5455,5357)(5481,5439)(5506,5516)
        (5530,5588)(5551,5656)(5572,5718)
        (5590,5775)(5607,5828)(5622,5875)
        (5635,5917)(5647,5954)(5657,5986)
        (5666,6014)(5674,6037)(5680,6056)
        (5685,6071)(5688,6083)(5691,6092)
        (5693,6098)(5694,6101)(5695,6103)(5695,6104)
\path(8095,3127)(8094,3127)(8092,3127)
        (8088,3127)(8081,3128)(8071,3128)
        (8058,3129)(8040,3130)(8019,3131)
        (7994,3132)(7963,3133)(7929,3135)
        (7889,3136)(7845,3138)(7796,3140)
        (7743,3142)(7685,3144)(7624,3146)
        (7559,3148)(7491,3150)(7420,3152)
        (7346,3154)(7271,3155)(7194,3157)
        (7116,3158)(7037,3159)(6958,3159)
        (6879,3159)(6801,3159)(6723,3158)
        (6646,3156)(6571,3154)(6497,3152)
        (6425,3148)(6354,3144)(6286,3140)
        (6220,3134)(6156,3127)(6095,3120)
        (6035,3112)(5978,3102)(5923,3092)
        (5870,3081)(5819,3068)(5769,3054)
        (5722,3039)(5676,3023)(5631,3005)
        (5588,2986)(5545,2965)(5503,2943)
        (5462,2919)(5421,2894)(5380,2866)
        (5340,2837)(5299,2805)(5258,2772)
        (5217,2736)(5175,2697)(5133,2657)
        (5091,2613)(5048,2568)(5004,2520)
        (4960,2470)(4915,2418)(4869,2363)
        (4823,2307)(4776,2249)(4729,2189)
        (4681,2128)(4634,2065)(4586,2002)
        (4538,1938)(4490,1873)(4443,1809)
        (4397,1745)(4351,1682)(4307,1620)
        (4264,1559)(4223,1500)(4183,1444)
        (4145,1390)(4110,1339)(4077,1291)
        (4046,1247)(4018,1206)(3992,1169)
        (3970,1136)(3950,1107)(3933,1082)
        (3918,1061)(3906,1043)(3896,1028)
        (3889,1017)(3883,1009)(3879,1003)
        (3877,1000)(3876,998)(3875,997)
\path(6375,1292)(6374,1293)(6373,1294)
        (6369,1297)(6364,1302)(6357,1308)
        (6347,1317)(6335,1329)(6319,1343)
        (6300,1361)(6278,1381)(6252,1405)
        (6223,1432)(6190,1461)(6154,1494)
        (6114,1530)(6072,1568)(6026,1609)
        (5978,1653)(5927,1698)(5873,1745)
        (5818,1794)(5761,1844)(5703,1894)
        (5644,1945)(5584,1997)(5524,2048)
        (5463,2099)(5403,2149)(5342,2198)
        (5282,2246)(5223,2293)(5165,2339)
        (5107,2383)(5051,2425)(4995,2465)
        (4941,2503)(4888,2539)(4835,2573)
        (4784,2605)(4734,2635)(4685,2663)
        (4637,2688)(4589,2711)(4542,2732)
        (4496,2752)(4449,2769)(4403,2784)
        (4357,2797)(4311,2809)(4264,2819)
        (4217,2828)(4169,2835)(4119,2840)
        (4069,2844)(4017,2846)(3963,2846)
        (3908,2845)(3851,2843)(3792,2839)
        (3730,2833)(3667,2826)(3602,2818)
        (3535,2809)(3466,2798)(3395,2786)
        (3322,2772)(3248,2758)(3172,2743)
        (3095,2727)(3017,2710)(2939,2693)
        (2860,2675)(2781,2656)(2702,2638)
        (2624,2619)(2548,2600)(2473,2581)
        (2400,2563)(2329,2545)(2261,2527)
        (2197,2510)(2136,2495)(2079,2479)
        (2026,2465)(1978,2453)(1934,2441)
        (1895,2430)(1860,2421)(1830,2413)
        (1805,2406)(1784,2400)(1767,2395)
        (1754,2392)(1744,2389)(1737,2387)
        (1733,2386)(1731,2385)(1730,2385)
\path(3873,997)(3873,998)(3873,1000)
        (3872,1004)(3871,1011)(3870,1021)
        (3868,1034)(3866,1051)(3863,1073)
        (3860,1098)(3856,1128)(3851,1163)
        (3846,1203)(3840,1247)(3833,1295)
        (3826,1348)(3818,1405)(3809,1466)
        (3800,1531)(3790,1599)(3779,1670)
        (3768,1742)(3756,1817)(3744,1894)
        (3732,1971)(3719,2049)(3705,2128)
        (3691,2206)(3677,2284)(3663,2361)
        (3648,2436)(3632,2510)(3617,2583)
        (3601,2654)(3585,2723)(3568,2789)
        (3551,2854)(3533,2916)(3515,2975)
        (3496,3033)(3477,3088)(3457,3140)
        (3436,3191)(3415,3239)(3392,3286)
        (3369,3330)(3345,3373)(3319,3414)
        (3293,3453)(3265,3492)(3236,3530)
        (3205,3566)(3173,3602)(3139,3638)
        (3102,3673)(3064,3707)(3023,3742)
        (2981,3776)(2935,3810)(2888,3845)
        (2837,3879)(2785,3914)(2730,3949)
        (2673,3984)(2613,4019)(2551,4055)
        (2487,4090)(2421,4126)(2354,4162)
        (2285,4199)(2215,4235)(2144,4271)
        (2072,4307)(2000,4343)(1928,4378)
        (1857,4413)(1786,4446)(1717,4479)
        (1649,4511)(1584,4542)(1521,4571)
        (1461,4599)(1405,4625)(1352,4650)
        (1302,4672)(1257,4693)(1216,4711)
        (1180,4728)(1147,4742)(1119,4755)
        (1095,4766)(1076,4774)(1060,4782)
        (1047,4787)(1038,4791)(1032,4794)
        (1028,4796)(1026,4797)(1025,4797)
\path(1730,2382)(1731,2383)(1732,2385)
        (1734,2388)(1738,2394)(1744,2402)
        (1751,2414)(1761,2428)(1772,2446)
        (1787,2468)(1804,2494)(1823,2523)
        (1845,2557)(1869,2595)(1896,2636)
        (1926,2682)(1957,2731)(1991,2783)
        (2027,2839)(2064,2897)(2102,2958)
        (2142,3021)(2182,3086)(2223,3152)
        (2265,3220)(2307,3288)(2348,3357)
        (2389,3426)(2429,3495)(2469,3563)
        (2507,3630)(2545,3697)(2581,3763)
        (2615,3827)(2648,3891)(2679,3953)
        (2708,4013)(2735,4072)(2761,4129)
        (2784,4185)(2806,4240)(2825,4293)
        (2842,4345)(2858,4396)(2871,4446)
        (2882,4495)(2891,4544)(2899,4592)
        (2904,4639)(2908,4687)(2910,4735)
        (2911,4783)(2909,4831)(2906,4881)
        (2901,4931)(2894,4982)(2885,5035)
        (2874,5089)(2862,5144)(2847,5202)
        (2831,5260)(2813,5321)(2793,5383)
        (2772,5447)(2748,5513)(2724,5580)
        (2698,5649)(2670,5719)(2641,5791)
        (2612,5863)(2581,5936)(2549,6010)
        (2517,6084)(2485,6158)(2452,6231)
        (2419,6304)(2386,6376)(2354,6446)
        (2323,6514)(2292,6580)(2263,6643)
        (2234,6703)(2207,6760)(2182,6813)
        (2159,6862)(2137,6907)(2117,6947)
        (2100,6984)(2084,7016)(2071,7044)
        (2059,7067)(2050,7087)(2042,7103)
        (2036,7115)(2031,7124)(2028,7130)
        (2026,7134)(2025,7136)(2025,7137)
\path(1025,4795)(1026,4795)(1028,4796)
        (1032,4797)(1039,4799)(1048,4802)
        (1061,4806)(1078,4811)(1098,4818)
        (1123,4826)(1152,4835)(1186,4846)
        (1224,4858)(1267,4872)(1314,4887)
        (1365,4903)(1421,4921)(1480,4941)
        (1542,4961)(1608,4983)(1676,5006)
        (1746,5030)(1819,5054)(1892,5080)
        (1967,5106)(2042,5132)(2117,5159)
        (2193,5187)(2267,5214)(2341,5242)
        (2413,5270)(2484,5298)(2553,5326)
        (2621,5354)(2686,5382)(2749,5410)
        (2810,5439)(2868,5467)(2924,5495)
        (2978,5523)(3029,5552)(3078,5581)
        (3124,5610)(3168,5639)(3210,5669)
        (3250,5700)(3288,5731)(3324,5763)
        (3359,5796)(3392,5830)(3424,5865)
        (3455,5901)(3485,5939)(3514,5979)
        (3542,6020)(3569,6064)(3596,6110)
        (3622,6157)(3648,6207)(3674,6260)
        (3699,6315)(3723,6372)(3748,6432)
        (3772,6494)(3797,6558)(3821,6625)
        (3845,6693)(3868,6764)(3892,6836)
        (3915,6909)(3939,6984)(3962,7060)
        (3984,7136)(4006,7213)(4028,7289)
        (4050,7364)(4071,7439)(4091,7512)
        (4110,7584)(4129,7653)(4146,7719)
        (4163,7782)(4179,7842)(4193,7898)
        (4206,7950)(4219,7997)(4230,8041)
        (4239,8079)(4248,8113)(4255,8143)
        (4262,8168)(4267,8189)(4271,8205)
        (4274,8218)(4277,8228)(4278,8235)
        (4279,8239)(4280,8241)(4280,8242)
\path(4288,8245)(4288,8244)(4289,8242)
        (4291,8238)(4294,8232)(4299,8223)
        (4305,8211)(4313,8196)(4323,8177)
        (4334,8154)(4348,8127)(4364,8096)
        (4382,8061)(4403,8021)(4425,7978)
        (4450,7930)(4477,7879)(4506,7825)
        (4536,7767)(4568,7706)(4602,7643)
        (4637,7578)(4673,7512)(4710,7444)
        (4748,7375)(4786,7306)(4825,7237)
        (4864,7168)(4904,7100)(4943,7032)
        (4982,6966)(5021,6901)(5060,6838)
        (5099,6777)(5137,6718)(5175,6661)
        (5213,6606)(5250,6554)(5287,6503)
        (5324,6456)(5360,6411)(5397,6368)
        (5433,6327)(5469,6289)(5505,6253)
        (5542,6219)(5579,6187)(5616,6157)
        (5654,6129)(5693,6102)(5733,6077)
        (5773,6052)(5816,6030)(5859,6008)
        (5905,5987)(5952,5968)(6001,5949)
        (6052,5931)(6106,5915)(6162,5898)
        (6220,5883)(6280,5868)(6343,5854)
        (6408,5841)(6475,5828)(6544,5816)
        (6615,5804)(6688,5792)(6763,5781)
        (6839,5771)(6916,5760)(6995,5751)
        (7073,5741)(7152,5732)(7230,5724)
        (7308,5715)(7385,5708)(7460,5700)
        (7533,5693)(7604,5686)(7672,5680)
        (7737,5675)(7798,5669)(7856,5664)
        (7909,5660)(7958,5656)(8002,5653)
        (8041,5649)(8076,5647)(8107,5644)
        (8132,5643)(8153,5641)(8171,5640)
        (8184,5639)(8194,5638)(8200,5638)
        (8205,5637)(8207,5637)(8208,5637)
\path(8210,5637)(8209,5636)(8207,5635)
        (8204,5632)(8199,5628)(8191,5622)
        (8180,5614)(8167,5604)(8150,5591)
        (8129,5575)(8105,5557)(8078,5535)
        (8046,5511)(8011,5484)(7973,5454)
        (7931,5421)(7886,5386)(7837,5348)
        (7786,5307)(7733,5265)(7677,5221)
        (7620,5175)(7561,5127)(7501,5079)
        (7441,5029)(7380,4979)(7319,4928)
        (7259,4877)(7199,4827)(7140,4776)
        (7082,4725)(7026,4675)(6971,4625)
        (6918,4576)(6867,4528)(6818,4480)
        (6771,4434)(6727,4387)(6684,4342)
        (6644,4297)(6607,4253)(6571,4210)
        (6538,4167)(6507,4125)(6479,4082)
        (6452,4040)(6428,3998)(6405,3956)
        (6384,3913)(6365,3870)(6347,3826)
        (6331,3782)(6317,3736)(6303,3689)
        (6291,3640)(6281,3590)(6272,3538)
        (6264,3484)(6257,3429)(6252,3371)
        (6247,3310)(6244,3248)(6242,3184)
        (6241,3117)(6240,3048)(6241,2978)
        (6243,2905)(6245,2831)(6248,2755)
        (6251,2678)(6256,2600)(6260,2521)
        (6266,2441)(6271,2362)(6277,2282)
        (6284,2204)(6290,2127)(6297,2051)
        (6303,1977)(6310,1906)(6317,1838)
        (6323,1772)(6329,1711)(6335,1653)
        (6340,1600)(6346,1551)(6350,1507)
        (6354,1467)(6358,1432)(6362,1402)
        (6364,1376)(6367,1355)(6369,1338)
        (6370,1324)(6371,1314)(6372,1308)
        (6373,1303)(6373,1301)(6373,1300)
\path(1350,7735)(1351,7735)(1352,7734)
        (1355,7733)(1360,7732)(1367,7730)
        (1376,7727)(1388,7724)(1404,7719)
        (1422,7713)(1444,7707)(1470,7699)
        (1500,7690)(1534,7679)(1573,7668)
        (1615,7655)(1662,7640)(1713,7625)
        (1768,7608)(1828,7590)(1891,7570)
        (1958,7549)(2030,7527)(2105,7504)
        (2183,7480)(2264,7454)(2348,7428)
        (2435,7400)(2525,7372)(2616,7343)
        (2710,7313)(2804,7283)(2901,7252)
        (2998,7220)(3095,7189)(3193,7156)
        (3292,7124)(3390,7091)(3488,7058)
        (3585,7025)(3682,6992)(3777,6959)
        (3872,6926)(3965,6893)(4057,6860)
        (4147,6827)(4236,6794)(4322,6762)
        (4408,6729)(4491,6697)(4572,6664)
        (4652,6632)(4729,6600)(4805,6568)
        (4879,6536)(4951,6504)(5021,6471)
        (5090,6439)(5156,6407)(5222,6374)
        (5285,6341)(5348,6308)(5408,6275)
        (5468,6241)(5527,6207)(5585,6172)
        (5642,6136)(5698,6100)(5754,6063)
        (5809,6025)(5865,5987)(5920,5947)
        (5975,5906)(6031,5863)(6086,5820)
        (6142,5775)(6199,5728)(6256,5680)
        (6313,5630)(6371,5579)(6430,5526)
        (6489,5471)(6550,5414)(6611,5356)
        (6673,5296)(6736,5234)(6800,5171)
        (6865,5106)(6930,5039)(6997,4971)
        (7064,4901)(7132,4830)(7200,4758)
        (7270,4684)(7339,4610)(7409,4534)
        (7479,4458)(7550,4381)(7620,4304)
        (7691,4227)(7760,4150)(7830,4073)
        (7899,3996)(7967,3920)(8034,3845)
        (8099,3772)(8163,3699)(8226,3628)
        (8287,3560)(8346,3493)(8403,3428)
        (8457,3366)(8510,3306)(8559,3250)
        (8606,3196)(8650,3146)(8691,3098)
        (8730,3054)(8765,3014)(8797,2977)
        (8827,2943)(8853,2912)(8876,2885)
        (8897,2861)(8915,2840)(8930,2823)
        (8943,2808)(8954,2796)(8962,2786)
        (8969,2778)(8973,2773)(8976,2769)
        (8978,2767)(8980,2766)(8980,2765)
\path(3678,82)(3678,83)(3677,84)
        (3677,87)(3676,92)(3674,99)
        (3672,109)(3669,121)(3665,137)
        (3661,156)(3655,178)(3649,205)
        (3642,235)(3634,270)(3625,309)
        (3615,352)(3603,400)(3591,452)
        (3578,508)(3564,569)(3549,634)
        (3534,703)(3517,776)(3500,852)
        (3481,932)(3463,1016)(3443,1102)
        (3423,1191)(3403,1283)(3382,1377)
        (3361,1473)(3340,1570)(3319,1669)
        (3297,1769)(3276,1870)(3254,1971)
        (3233,2073)(3212,2174)(3192,2276)
        (3172,2377)(3152,2477)(3133,2577)
        (3114,2675)(3096,2773)(3078,2869)
        (3061,2963)(3045,3057)(3030,3148)
        (3016,3239)(3002,3327)(2989,3414)
        (2977,3499)(2966,3582)(2956,3664)
        (2946,3744)(2938,3823)(2931,3900)
        (2924,3975)(2919,4049)(2914,4122)
        (2911,4194)(2908,4264)(2907,4334)
        (2906,4402)(2906,4470)(2908,4538)
        (2910,4605)(2913,4672)(2917,4739)
        (2922,4806)(2928,4873)(2934,4941)
        (2942,5009)(2951,5078)(2961,5148)
        (2972,5219)(2984,5291)(2997,5365)
        (3012,5439)(3027,5515)(3043,5593)
        (3061,5672)(3080,5752)(3099,5834)
        (3120,5918)(3142,6003)(3165,6090)
        (3189,6179)(3214,6269)(3240,6360)
        (3266,6453)(3294,6548)(3322,6643)
        (3351,6740)(3381,6837)(3411,6935)
        (3441,7034)(3473,7133)(3504,7233)
        (3536,7332)(3567,7431)(3599,7529)
        (3631,7627)(3663,7724)(3694,7819)
        (3725,7913)(3755,8004)(3785,8094)
        (3814,8181)(3843,8265)(3870,8347)
        (3897,8425)(3922,8500)(3946,8571)
        (3969,8638)(3991,8702)(4011,8761)
        (4030,8816)(4047,8867)(4064,8913)
        (4078,8956)(4091,8994)(4103,9027)
        (4113,9057)(4122,9083)(4130,9105)
        (4136,9124)(4142,9139)(4146,9151)
        (4149,9160)(4152,9167)(4153,9172)
        (4154,9175)(4155,9176)(4155,9177)
\path(6820,462)(6819,462)(6818,463)
        (6816,465)(6812,468)(6806,472)
        (6798,478)(6788,486)(6775,496)
        (6759,507)(6740,521)(6718,537)
        (6693,556)(6664,578)(6632,602)
        (6596,628)(6557,658)(6514,690)
        (6468,725)(6418,762)(6364,802)
        (6307,845)(6247,890)(6185,938)
        (6119,987)(6050,1039)(5980,1093)
        (5907,1149)(5832,1206)(5755,1264)
        (5677,1324)(5598,1385)(5517,1447)
        (5436,1510)(5355,1574)(5273,1638)
        (5191,1702)(5109,1766)(5028,1831)
        (4947,1895)(4867,1959)(4788,2023)
        (4710,2087)(4633,2150)(4557,2212)
        (4483,2274)(4411,2335)(4340,2395)
        (4270,2455)(4203,2514)(4137,2572)
        (4073,2630)(4010,2687)(3950,2743)
        (3891,2798)(3834,2853)(3779,2907)
        (3725,2961)(3673,3014)(3623,3067)
        (3574,3119)(3527,3172)(3481,3224)
        (3436,3276)(3393,3329)(3350,3381)
        (3309,3434)(3268,3487)(3228,3541)
        (3189,3595)(3150,3650)(3112,3707)
        (3074,3764)(3036,3823)(2999,3882)
        (2962,3944)(2925,4007)(2889,4071)
        (2852,4138)(2815,4206)(2778,4275)
        (2741,4347)(2704,4421)(2667,4496)
        (2629,4574)(2591,4653)(2554,4734)
        (2515,4817)(2477,4902)(2438,4988)
        (2399,5077)(2360,5166)(2321,5258)
        (2282,5350)(2242,5443)(2203,5538)
        (2163,5633)(2124,5729)(2085,5825)
        (2046,5921)(2007,6018)(1969,6113)
        (1931,6208)(1893,6303)(1857,6396)
        (1821,6487)(1786,6577)(1751,6665)
        (1718,6750)(1686,6833)(1655,6912)
        (1626,6989)(1598,7063)(1571,7133)
        (1546,7199)(1522,7261)(1500,7319)
        (1479,7374)(1460,7424)(1443,7470)
        (1427,7511)(1413,7549)(1400,7582)
        (1389,7612)(1380,7637)(1372,7659)
        (1365,7677)(1359,7692)(1355,7704)
        (1351,7714)(1349,7720)(1347,7725)
        (1346,7728)(1345,7729)(1345,7730)
\path(8978,2770)(8977,2770)(8976,2770)
        (8973,2770)(8968,2769)(8960,2769)
        (8951,2768)(8938,2768)(8922,2767)
        (8902,2766)(8879,2764)(8852,2763)
        (8820,2761)(8784,2759)(8744,2757)
        (8700,2755)(8651,2752)(8597,2749)
        (8539,2746)(8477,2743)(8410,2739)
        (8339,2736)(8264,2732)(8185,2728)
        (8103,2725)(8017,2721)(7929,2716)
        (7837,2712)(7743,2708)(7646,2704)
        (7548,2700)(7448,2696)(7347,2693)
        (7244,2689)(7141,2685)(7037,2682)
        (6933,2679)(6829,2676)(6725,2673)
        (6622,2671)(6519,2669)(6418,2667)
        (6317,2666)(6217,2665)(6119,2665)
        (6023,2665)(5928,2665)(5835,2666)
        (5743,2668)(5653,2670)(5565,2672)
        (5479,2675)(5395,2679)(5312,2683)
        (5232,2688)(5153,2693)(5075,2699)
        (5000,2706)(4926,2714)(4853,2722)
        (4782,2731)(4712,2741)(4643,2752)
        (4575,2763)(4508,2775)(4441,2788)
        (4375,2802)(4310,2817)(4245,2833)
        (4179,2849)(4114,2867)(4048,2885)
        (3982,2905)(3916,2926)(3849,2948)
        (3781,2971)(3712,2996)(3642,3021)
        (3571,3049)(3498,3077)(3425,3107)
        (3350,3138)(3274,3171)(3197,3205)
        (3118,3240)(3038,3276)(2956,3314)
        (2873,3353)(2789,3394)(2703,3435)
        (2616,3478)(2528,3521)(2439,3566)
        (2349,3611)(2258,3658)(2167,3705)
        (2075,3752)(1983,3800)(1891,3849)
        (1798,3897)(1706,3946)(1615,3995)
        (1524,4043)(1435,4091)(1346,4139)
        (1260,4186)(1175,4232)(1092,4277)
        (1011,4321)(933,4364)(857,4405)
        (785,4445)(716,4483)(650,4520)
        (588,4554)(529,4587)(475,4617)
        (424,4645)(377,4672)(334,4696)
        (295,4717)(259,4737)(228,4754)
        (201,4770)(177,4783)(156,4795)
        (139,4804)(125,4812)(114,4819)
        (105,4823)(99,4827)(95,4829)
        (92,4831)(91,4832)(90,4832)
\path(9120,5932)(9119,5932)(9118,5930)
        (9116,5928)(9113,5925)(9107,5920)
        (9100,5913)(9091,5904)(9079,5893)
        (9065,5880)(9049,5864)(9029,5845)
        (9006,5824)(8980,5799)(8951,5772)
        (8919,5741)(8884,5707)(8845,5671)
        (8803,5631)(8757,5588)(8709,5543)
        (8658,5494)(8603,5443)(8546,5390)
        (8486,5334)(8424,5276)(8359,5215)
        (8292,5153)(8223,5089)(8153,5024)
        (8081,4958)(8007,4890)(7933,4822)
        (7857,4753)(7781,4684)(7705,4615)
        (7628,4545)(7551,4476)(7474,4407)
        (7397,4339)(7320,4272)(7244,4205)
        (7169,4139)(7094,4075)(7020,4011)
        (6946,3949)(6874,3888)(6803,3829)
        (6732,3771)(6663,3715)(6594,3660)
        (6527,3607)(6460,3556)(6395,3506)
        (6330,3458)(6267,3411)(6204,3366)
        (6142,3323)(6081,3281)(6020,3241)
        (5960,3202)(5900,3164)(5841,3128)
        (5782,3094)(5723,3060)(5664,3027)
        (5604,2996)(5545,2965)(5485,2935)
        (5425,2906)(5363,2877)(5301,2850)
        (5238,2822)(5174,2796)(5108,2770)
        (5041,2744)(4972,2719)(4902,2694)
        (4830,2669)(4756,2645)(4681,2621)
        (4603,2597)(4524,2573)(4443,2550)
        (4359,2526)(4274,2503)(4187,2480)
        (4098,2457)(4007,2434)(3915,2411)
        (3821,2388)(3725,2365)(3628,2343)
        (3529,2320)(3430,2298)(3329,2275)
        (3228,2253)(3126,2231)(3024,2209)
        (2921,2188)(2819,2166)(2717,2145)
        (2616,2125)(2516,2104)(2418,2084)
        (2321,2065)(2225,2046)(2132,2028)
        (2042,2010)(1954,1993)(1870,1977)
        (1788,1961)(1711,1946)(1636,1932)
        (1566,1919)(1500,1906)(1439,1895)
        (1381,1884)(1328,1874)(1280,1865)
        (1235,1857)(1196,1849)(1160,1843)
        (1129,1837)(1102,1832)(1079,1828)
        (1060,1824)(1044,1821)(1031,1819)
        (1021,1817)(1014,1816)(1009,1815)
        (1006,1814)(1005,1814)(1004,1814)
\path(7215,8457)(7215,8456)(7215,8455)
        (7214,8452)(7214,8447)(7213,8440)
        (7212,8430)(7210,8417)(7208,8401)
        (7206,8382)(7203,8359)(7200,8332)
        (7196,8300)(7192,8265)(7187,8225)
        (7182,8181)(7176,8132)(7169,8079)
        (7162,8021)(7155,7959)(7146,7893)
        (7138,7823)(7128,7749)(7118,7671)
        (7108,7589)(7097,7504)(7086,7416)
        (7074,7325)(7062,7232)(7049,7137)
        (7036,7039)(7022,6940)(7008,6840)
        (6994,6738)(6980,6636)(6965,6534)
        (6950,6431)(6935,6328)(6920,6226)
        (6904,6124)(6888,6023)(6872,5922)
        (6856,5823)(6840,5725)(6823,5629)
        (6807,5534)(6790,5441)(6773,5349)
        (6756,5259)(6738,5171)(6721,5085)
        (6703,5001)(6685,4919)(6666,4839)
        (6648,4760)(6629,4684)(6609,4609)
        (6590,4536)(6569,4464)(6549,4394)
        (6528,4326)(6506,4258)(6484,4193)
        (6461,4128)(6438,4064)(6413,4001)
        (6389,3939)(6363,3877)(6337,3816)
        (6309,3754)(6281,3693)(6252,3632)
        (6221,3570)(6189,3509)(6156,3446)
        (6122,3384)(6086,3320)(6049,3256)
        (6010,3191)(5970,3125)(5929,3058)
        (5886,2990)(5841,2921)(5794,2851)
        (5747,2779)(5697,2707)(5647,2633)
        (5594,2559)(5541,2483)(5485,2406)
        (5429,2328)(5372,2249)(5313,2169)
        (5253,2089)(5193,2008)(5131,1926)
        (5069,1844)(5007,1762)(4944,1680)
        (4880,1597)(4817,1516)(4754,1434)
        (4691,1354)(4629,1274)(4568,1196)
        (4507,1118)(4448,1043)(4389,969)
        (4333,898)(4278,828)(4225,762)
        (4173,697)(4125,636)(4078,578)
        (4034,523)(3992,471)(3953,422)
        (3917,377)(3883,336)(3853,298)
        (3825,263)(3800,232)(3777,204)
        (3757,180)(3740,159)(3726,141)
        (3713,125)(3703,113)(3695,103)
        (3689,96)(3684,90)(3681,86)
        (3679,84)(3678,83)(3678,82)
\path(2674,6580)(2673,6579)(2671,6577)
        (2667,6572)(2661,6565)(2652,6555)
        (2640,6541)(2625,6523)(2607,6502)
        (2585,6475)(2559,6445)(2530,6410)
        (2497,6371)(2462,6328)(2423,6281)
        (2382,6230)(2338,6175)(2293,6118)
        (2247,6059)(2199,5997)(2152,5935)
        (2104,5870)(2057,5806)(2011,5741)
        (1967,5677)(1925,5613)(1885,5550)
        (1848,5488)(1815,5428)(1784,5370)
        (1758,5313)(1735,5259)(1717,5207)
        (1703,5157)(1694,5109)(1690,5064)
        (1691,5020)(1697,4979)(1708,4940)
        (1725,4903)(1747,4867)(1775,4834)
        (1809,4801)(1848,4770)(1877,4750)
        (1909,4730)(1944,4710)(1981,4691)
        (2022,4672)(2066,4653)(2113,4635)
        (2163,4616)(2217,4598)(2274,4580)
        (2335,4562)(2400,4544)(2468,4526)
        (2539,4508)(2614,4490)(2693,4472)
        (2775,4454)(2860,4435)(2949,4417)
        (3041,4399)(3135,4380)(3233,4362)
        (3334,4343)(3437,4325)(3543,4306)
        (3651,4287)(3760,4269)(3872,4250)
        (3984,4232)(4098,4213)(4212,4195)
        (4327,4177)(4442,4159)(4556,4141)
        (4669,4123)(4782,4106)(4892,4090)
        (5001,4073)(5108,4057)(5211,4042)
        (5312,4027)(5409,4013)(5502,4000)
        (5591,3987)(5675,3974)(5755,3963)
        (5830,3952)(5900,3942)(5965,3933)
        (6024,3925)(6078,3917)(6127,3910)
        (6171,3904)(6209,3899)(6243,3894)
        (6271,3890)(6296,3887)(6315,3884)
        (6331,3882)(6343,3880)(6352,3879)
        (6358,3878)(6362,3877)(6364,3877)(6365,3877)
\path(4393,7445)(4392,7445)(4389,7444)
        (4383,7443)(4374,7442)(4361,7440)
        (4344,7437)(4322,7433)(4294,7428)
        (4262,7422)(4224,7415)(4180,7407)
        (4132,7397)(4078,7387)(4020,7375)
        (3958,7363)(3892,7349)(3823,7334)
        (3752,7318)(3678,7301)(3604,7283)
        (3529,7264)(3454,7244)(3380,7224)
        (3307,7202)(3236,7180)(3168,7157)
        (3102,7133)(3040,7109)(2982,7083)
        (2927,7056)(2877,7029)(2831,7000)
        (2791,6971)(2755,6939)(2724,6907)
        (2698,6873)(2678,6837)(2662,6799)
        (2653,6760)(2648,6718)(2649,6673)
        (2654,6627)(2665,6577)(2675,6542)
        (2687,6506)(2701,6469)(2718,6430)
        (2737,6389)(2759,6346)(2784,6301)
        (2811,6254)(2840,6205)(2873,6154)
        (2908,6100)(2946,6045)(2986,5987)
        (3030,5927)(3076,5864)(3124,5799)
        (3175,5732)(3229,5663)(3285,5591)
        (3344,5518)(3404,5442)(3467,5364)
        (3533,5285)(3600,5203)(3668,5121)
        (3739,5037)(3811,4951)(3884,4865)
        (3958,4777)(4033,4690)(4108,4601)
        (4184,4513)(4260,4425)(4336,4337)
        (4412,4250)(4487,4164)(4560,4080)
        (4633,3997)(4704,3915)(4773,3837)
        (4840,3760)(4905,3686)(4968,3616)
        (5027,3548)(5084,3484)(5138,3423)
        (5188,3366)(5235,3313)(5279,3264)
        (5319,3219)(5355,3178)(5388,3141)
        (5417,3109)(5443,3079)(5466,3054)
        (5485,3033)(5501,3014)(5514,2999)
        (5525,2988)(5533,2978)(5539,2972)
        (5543,2967)(5546,2964)(5547,2963)(5548,2962)
\path(1000,1815)(1000,1816)(1001,1817)
        (1002,1820)(1005,1824)(1008,1831)
        (1012,1839)(1018,1851)(1025,1865)
        (1034,1883)(1045,1903)(1057,1928)
        (1071,1956)(1088,1988)(1106,2023)
        (1126,2063)(1149,2107)(1173,2154)
        (1200,2206)(1228,2261)(1259,2321)
        (1291,2384)(1326,2450)(1362,2520)
        (1400,2593)(1439,2669)(1480,2748)
        (1523,2829)(1566,2912)(1611,2998)
        (1657,3085)(1704,3173)(1752,3262)
        (1800,3353)(1849,3444)(1898,3535)
        (1947,3626)(1997,3718)(2047,3809)
        (2096,3899)(2146,3989)(2196,4077)
        (2245,4165)(2294,4251)(2343,4336)
        (2391,4419)(2439,4501)(2487,4581)
        (2534,4659)(2581,4736)(2627,4811)
        (2673,4884)(2718,4955)(2763,5024)
        (2808,5091)(2852,5156)(2896,5220)
        (2940,5282)(2983,5342)(3027,5401)
        (3070,5458)(3114,5513)(3157,5567)
        (3201,5620)(3245,5672)(3290,5723)
        (3335,5773)(3380,5822)(3426,5871)
        (3473,5919)(3521,5966)(3569,6014)
        (3619,6061)(3671,6108)(3723,6155)
        (3777,6202)(3833,6249)(3890,6297)
        (3949,6344)(4009,6393)(4071,6441)
        (4136,6490)(4202,6539)(4269,6589)
        (4339,6640)(4410,6691)(4484,6742)
        (4559,6794)(4635,6847)(4714,6900)
        (4794,6954)(4875,7008)(4958,7062)
        (5042,7117)(5127,7173)(5213,7228)
        (5300,7284)(5387,7339)(5475,7395)
        (5563,7450)(5650,7505)(5738,7559)
        (5825,7613)(5911,7667)(5996,7719)
        (6079,7771)(6162,7821)(6242,7870)
        (6320,7917)(6396,7964)(6469,8008)
        (6539,8050)(6606,8091)(6670,8130)
        (6731,8166)(6788,8201)(6841,8233)
        (6891,8262)(6937,8290)(6979,8315)
        (7017,8338)(7052,8358)(7082,8377)
        (7109,8393)(7133,8407)(7153,8418)
        (7170,8428)(7183,8437)(7194,8443)
        (7203,8448)(7209,8452)(7213,8454)
        (7216,8456)(7217,8457)(7218,8457)
\path(2030,7137)(2031,7137)(2033,7136)
        (2037,7134)(2043,7131)(2053,7127)
        (2065,7122)(2081,7115)(2101,7107)
        (2125,7096)(2153,7085)(2186,7071)
        (2223,7055)(2264,7038)(2310,7019)
        (2360,6998)(2414,6975)(2471,6952)
        (2532,6927)(2596,6901)(2663,6874)
        (2732,6846)(2803,6817)(2876,6789)
        (2950,6760)(3024,6731)(3099,6703)
        (3174,6675)(3249,6647)(3323,6620)
        (3397,6595)(3469,6570)(3540,6546)
        (3610,6524)(3677,6503)(3744,6483)
        (3808,6465)(3871,6448)(3932,6434)
        (3991,6420)(4049,6409)(4104,6399)
        (4158,6391)(4211,6385)(4262,6381)
        (4312,6378)(4361,6378)(4409,6379)
        (4457,6382)(4504,6386)(4551,6392)
        (4598,6400)(4645,6410)(4692,6421)
        (4740,6435)(4789,6450)(4839,6468)
        (4889,6488)(4941,6510)(4994,6534)
        (5048,6560)(5104,6588)(5161,6618)
        (5219,6650)(5279,6683)(5339,6719)
        (5401,6756)(5465,6795)(5529,6835)
        (5594,6877)(5659,6919)(5725,6963)
        (5791,7007)(5857,7051)(5922,7096)
        (5987,7140)(6050,7184)(6112,7228)
        (6173,7270)(6231,7311)(6287,7351)
        (6340,7389)(6390,7425)(6437,7458)
        (6480,7490)(6520,7518)(6556,7544)
        (6588,7568)(6616,7589)(6641,7606)
        (6662,7622)(6679,7634)(6693,7645)
        (6703,7653)(6711,7658)(6717,7663)
        (6720,7665)(6722,7666)(6723,7667)
\path(90,4830)(91,4830)(92,4831)
        (95,4832)(100,4834)(106,4837)
        (115,4841)(127,4846)(142,4852)
        (160,4860)(181,4869)(206,4880)
        (235,4892)(268,4906)(305,4922)
        (347,4940)(392,4959)(441,4980)
        (495,5003)(553,5027)(614,5053)
        (680,5081)(749,5110)(822,5140)
        (898,5172)(977,5205)(1059,5239)
        (1144,5274)(1231,5310)(1321,5347)
        (1412,5385)(1505,5422)(1599,5461)
        (1694,5499)(1790,5538)(1887,5577)
        (1983,5615)(2080,5653)(2177,5691)
        (2273,5729)(2369,5766)(2464,5802)
        (2558,5838)(2652,5872)(2744,5906)
        (2834,5939)(2924,5971)(3012,6003)
        (3099,6032)(3184,6061)(3267,6089)
        (3349,6116)(3430,6141)(3509,6165)
        (3586,6188)(3662,6210)(3737,6231)
        (3810,6250)(3882,6268)(3953,6285)
        (4023,6301)(4092,6316)(4161,6330)
        (4228,6342)(4295,6354)(4362,6364)
        (4429,6373)(4495,6382)(4561,6390)
        (4628,6396)(4695,6402)(4763,6407)
        (4832,6411)(4901,6415)(4972,6417)
        (5043,6418)(5116,6419)(5190,6418)
        (5266,6417)(5343,6415)(5422,6412)
        (5503,6408)(5585,6404)(5669,6398)
        (5754,6392)(5842,6385)(5931,6377)
        (6022,6369)(6114,6360)(6208,6350)
        (6304,6340)(6401,6329)(6499,6317)
        (6599,6305)(6699,6293)(6801,6280)
        (6903,6267)(7005,6253)(7108,6239)
        (7211,6225)(7313,6210)(7415,6196)
        (7516,6181)(7616,6167)(7715,6152)
        (7812,6138)(7907,6123)(8000,6109)
        (8090,6095)(8177,6082)(8262,6069)
        (8343,6056)(8420,6044)(8494,6032)
        (8564,6021)(8630,6011)(8691,6001)
        (8748,5991)(8801,5983)(8849,5975)
        (8893,5968)(8933,5961)(8968,5956)
        (8999,5951)(9026,5946)(9048,5942)
        (9068,5939)(9084,5937)(9096,5934)
        (9106,5933)(9113,5932)(9118,5931)
        (9121,5930)(9122,5930)(9123,5930)
\path(6293,6987)(6292,6988)(6290,6989)
        (6285,6992)(6278,6996)(6268,7002)
        (6254,7010)(6237,7020)(6215,7032)
        (6189,7047)(6159,7064)(6124,7084)
        (6084,7106)(6041,7130)(5993,7157)
        (5941,7185)(5886,7214)(5828,7246)
        (5767,7278)(5703,7310)(5638,7344)
        (5572,7377)(5504,7409)(5436,7442)
        (5368,7473)(5301,7502)(5234,7530)
        (5169,7556)(5105,7580)(5043,7602)
        (4983,7620)(4925,7636)(4870,7648)
        (4817,7657)(4767,7663)(4720,7665)
        (4675,7662)(4633,7656)(4594,7646)
        (4557,7631)(4524,7612)(4492,7588)
        (4463,7559)(4437,7526)(4412,7488)
        (4390,7445)(4376,7413)(4362,7379)
        (4350,7342)(4338,7303)(4327,7261)
        (4316,7216)(4307,7168)(4298,7117)
        (4289,7062)(4282,7005)(4275,6944)
        (4268,6880)(4263,6813)(4257,6742)
        (4253,6668)(4249,6591)(4245,6511)
        (4242,6427)(4240,6341)(4238,6251)
        (4236,6158)(4235,6063)(4234,5965)
        (4234,5864)(4234,5761)(4234,5656)
        (4235,5548)(4236,5439)(4237,5329)
        (4239,5217)(4240,5105)(4242,4992)
        (4245,4879)(4247,4765)(4250,4652)
        (4252,4540)(4255,4429)(4258,4320)
        (4261,4212)(4264,4107)(4267,4004)
        (4270,3905)(4273,3808)(4276,3715)
        (4279,3626)(4282,3542)(4284,3461)
        (4287,3385)(4289,3314)(4292,3248)
        (4294,3187)(4296,3131)(4298,3080)
        (4300,3034)(4301,2993)(4302,2957)
        (4304,2926)(4305,2899)(4305,2877)
        (4306,2858)(4307,2844)(4307,2832)
        (4308,2824)(4308,2818)(4308,2815)
        (4308,2813)(4308,2812)
\path(4045,1795)(4046,1795)(4049,1794)
        (4054,1793)(4062,1792)(4074,1790)
        (4089,1787)(4109,1783)(4134,1779)
        (4163,1774)(4198,1768)(4237,1761)
        (4282,1754)(4331,1746)(4385,1737)
        (4444,1728)(4506,1719)(4571,1709)
        (4640,1699)(4711,1690)(4784,1681)
        (4858,1672)(4932,1664)(5007,1657)
        (5082,1651)(5155,1646)(5228,1642)
        (5298,1639)(5367,1638)(5432,1639)
        (5495,1642)(5555,1647)(5611,1654)
        (5664,1663)(5714,1675)(5759,1689)
        (5800,1706)(5838,1726)(5871,1749)
        (5900,1776)(5926,1805)(5947,1838)
        (5965,1875)(5978,1915)(5988,1959)
        (5995,2007)(5998,2042)(5999,2078)
        (5998,2117)(5996,2158)(5992,2202)
        (5986,2247)(5979,2296)(5970,2347)
        (5960,2401)(5947,2457)(5933,2517)
        (5918,2579)(5900,2644)(5881,2712)
        (5860,2783)(5838,2857)(5814,2934)
        (5788,3013)(5761,3096)(5733,3181)
        (5703,3268)(5671,3359)(5639,3451)
        (5605,3546)(5570,3643)(5534,3742)
        (5497,3842)(5459,3944)(5420,4047)
        (5380,4152)(5340,4257)(5300,4362)
        (5259,4468)(5219,4574)(5178,4679)
        (5137,4783)(5097,4887)(5057,4989)
        (5018,5089)(4979,5187)(4941,5282)
        (4904,5375)(4869,5464)(4834,5551)
        (4801,5633)(4770,5712)(4740,5787)
        (4712,5857)(4685,5923)(4661,5984)
        (4638,6041)(4617,6093)(4598,6140)
        (4581,6183)(4565,6221)(4552,6254)
        (4540,6283)(4530,6308)(4521,6329)
        (4514,6346)(4509,6360)(4505,6370)
        (4501,6378)(4499,6383)(4498,6387)
        (4497,6388)(4497,6389)
\put(6815,470){\whiten\ellipse{160}{160}}
\put(6815,470){\ellipse{160}{160}}
\put(6368,1305){\whiten\ellipse{160}{160}}
\put(6368,1305){\ellipse{160}{160}}
\put(5998,2007){\whiten\ellipse{160}{160}}
\put(5998,2007){\ellipse{160}{160}}
\put(7350,3427){\whiten\ellipse{160}{160}}
\put(7350,3427){\ellipse{160}{160}}
\put(8975,2772){\whiten\ellipse{160}{160}}
\put(8975,2772){\ellipse{160}{160}}
\put(8088,3130){\whiten\ellipse{160}{160}}
\put(8088,3130){\ellipse{160}{160}}
\put(9120,5932){\whiten\ellipse{160}{160}}
\put(9120,5932){\ellipse{160}{160}}
\put(8210,5640){\whiten\ellipse{160}{160}}
\put(8210,5640){\ellipse{160}{160}}
\put(7453,5392){\whiten\ellipse{160}{160}}
\put(7453,5392){\ellipse{160}{160}}
\put(6293,6987){\whiten\ellipse{160}{160}}
\put(6293,6987){\ellipse{160}{160}}
\put(6718,7662){\whiten\ellipse{160}{160}}
\put(6718,7662){\ellipse{160}{160}}
\put(7215,8455){\whiten\ellipse{160}{160}}
\put(7215,8455){\ellipse{160}{160}}
\put(4393,7450){\whiten\ellipse{160}{160}}
\put(4393,7450){\ellipse{160}{160}}
\put(4283,8237){\whiten\ellipse{160}{160}}
\put(4283,8237){\ellipse{160}{160}}
\put(4155,9177){\whiten\ellipse{160}{160}}
\put(4155,9177){\ellipse{160}{160}}
\put(2025,7137){\whiten\ellipse{160}{160}}
\put(2025,7137){\ellipse{160}{160}}
\put(2668,6577){\whiten\ellipse{160}{160}}
\put(2668,6577){\ellipse{160}{160}}
\put(1348,7735){\whiten\ellipse{160}{160}}
\put(1348,7735){\ellipse{160}{160}}
\put(1018,4797){\whiten\ellipse{160}{160}}
\put(1018,4797){\ellipse{160}{160}}
\put(1853,4767){\whiten\ellipse{160}{160}}
\put(1853,4767){\ellipse{160}{160}}
\put(88,4832){\whiten\ellipse{160}{160}}
\put(88,4832){\ellipse{160}{160}}
\put(1733,2385){\whiten\ellipse{160}{160}}
\put(1733,2385){\ellipse{160}{160}}
\put(2390,2892){\whiten\ellipse{160}{160}}
\put(2390,2892){\ellipse{160}{160}}
\put(1000,1815){\whiten\ellipse{160}{160}}
\put(1000,1815){\ellipse{160}{160}}
\put(4045,1795){\whiten\ellipse{160}{160}}
\put(4045,1795){\ellipse{160}{160}}
\put(3878,1005){\whiten\ellipse{160}{160}}
\put(3878,1005){\ellipse{160}{160}}
\put(3683,87){\whiten\ellipse{160}{160}}
\put(3683,87){\ellipse{160}{160}}
\put(4305,2815){\whiten\ellipse{160}{160}}
\put(4305,2815){\ellipse{160}{160}}
\put(5543,2965){\whiten\ellipse{160}{160}}
\put(5543,2965){\ellipse{160}{160}}
\put(3265,3487){\whiten\ellipse{160}{160}}
\put(3265,3487){\ellipse{160}{160}}
\put(6368,3877){\whiten\ellipse{160}{160}}
\put(6368,3877){\ellipse{160}{160}}
\put(6438,5100){\whiten\ellipse{160}{160}}
\put(6438,5100){\ellipse{160}{160}}
\put(5695,6105){\whiten\ellipse{160}{160}}
\put(5695,6105){\ellipse{160}{160}}
\put(4500,6382){\whiten\ellipse{160}{160}}
\put(4500,6382){\ellipse{160}{160}}
\put(3385,5822){\whiten\ellipse{160}{160}}
\put(3385,5822){\ellipse{160}{160}}
\put(2910,4680){\whiten\ellipse{160}{160}}
\put(2910,4680){\ellipse{160}{160}}
\end{picture}
}
\caption{The unique Greechie-3-L diagram of type $(36,36)$\label{g36}}
\end{figure}
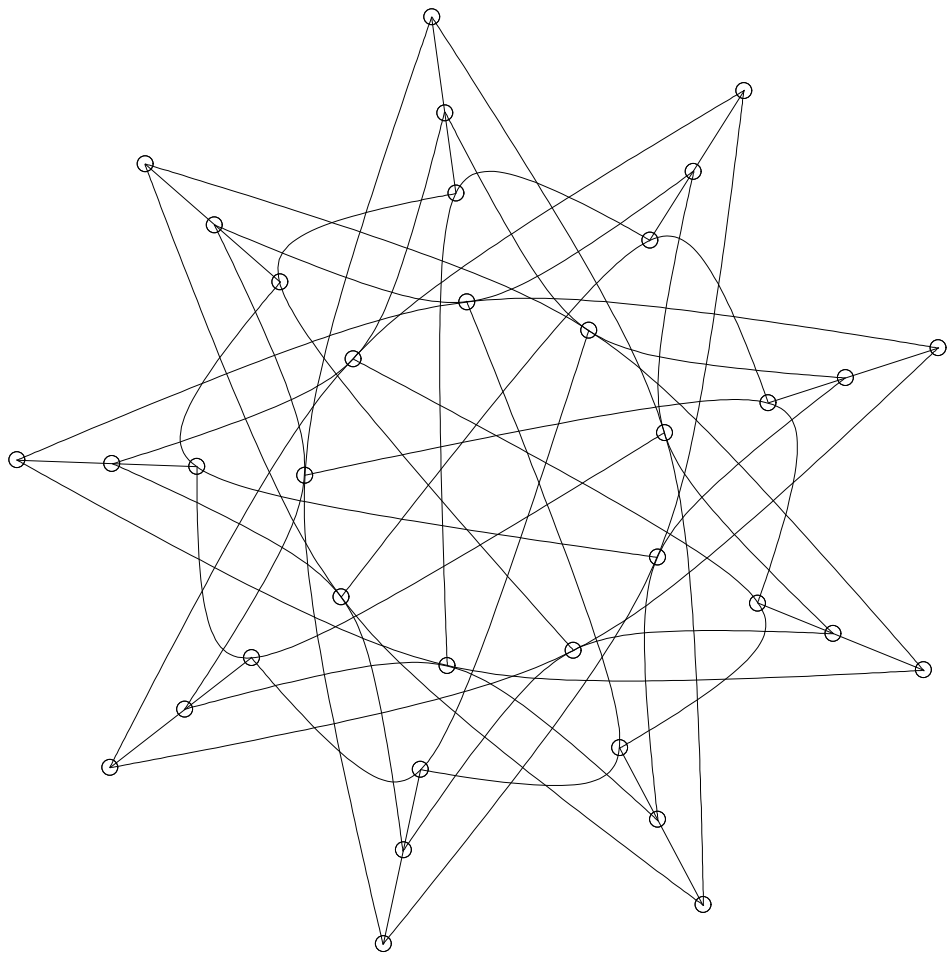

%----------------------------------------------------------------

Of some interest is that the Greechie-3-L diagrams of type $(35,35)$ and
$(36,36)$ contain only atoms of rank~3.  There is a connection here to
known results in graph theory, as follows.  Let $D$ be a diagram in one
of these two classes.  Define a graph $G$ whose vertices are the atoms
and blocks of~$D$.  A vertex which is an atom is adjacent in $G$ to a
vertex which is a block provided the atom lies in the block, and there
are no other edges.  (This is the same incidence graph we defined earlier.)
Then each vertex of the graph has valence~3, and
moreover there are no cycles of length~9 or less (which is equivalent
to the requirement that $D$ has no loops of length~4 or less).
In graph theoretic language, $G$ is a bipartite cubic (or trivalent)
graph of girth at least~10.  The smallest such graphs are three with
70 vertices \cite{OKeefe}.  In two of the three graphs, the functions of
``atom'' and ``block'' can be interchanged to produce different Greechie
diagrams, but in the third this interchange gives an isomorphic diagram.
That is how the three graphs correspond to the five diagrams we found.

Figure~\ref{g36} shows the unique Greechie-3-L diagram of type $(36,36)$.
Using a slight adaptation of the program described in \cite{mmn98}, we
have found that there are no Greechie-3-L diagrams of
type $(37,37)$ with every atom having rank~3, and exactly eight such
diagrams of type $(38,38)$.

\section{Testing conjectures with Greechie diagrams}
\label{sec:latticeg}

In order to test equations conjectured to hold in various classes of
orthomodular lattices, it is useful to automate their checking against
Greechie diagrams.  This will let us either falsify the conjecture or
give us some confidence that might hold in the class of interest before
we attempt to prove it.  Also, finding a lattice in which one equation
holds but a second one fails gives us a proof that the second is
independent.

To this end we use a program called {\tt latticeg}\footnote{Available at
{\tt ftp://ftp.shore.net/members/ndm/quantum-logic} as the {\sc ansi} C
program {\tt latticeg.c}.  The program is simple to use and
self-explanatory with the {\tt --help} option.  A related program {\tt
lattice.c} handles general Hasse diagrams and has built-in those
lattices we have found most useful for preliminary testing of
conjectures.  These two programs, along with {\tt beran.c} for computing
the canonical orthomodular form of any two-variable expression, are the
primary computational tools we have used for studying orthomodular and
Hilbert lattice equations.} which will check to see if an equation
or inference holds in each of the Greechie diagrams in a list provided
by the user.

The {\tt latticeg} program internally converts a Greechie diagram to
its corresponding Hasse diagram and tests all possible assignments of nodes
in the Hasse diagram to the equation.
In general Greechie diagrams correspond to Boolean algebras ``pasted''
together.

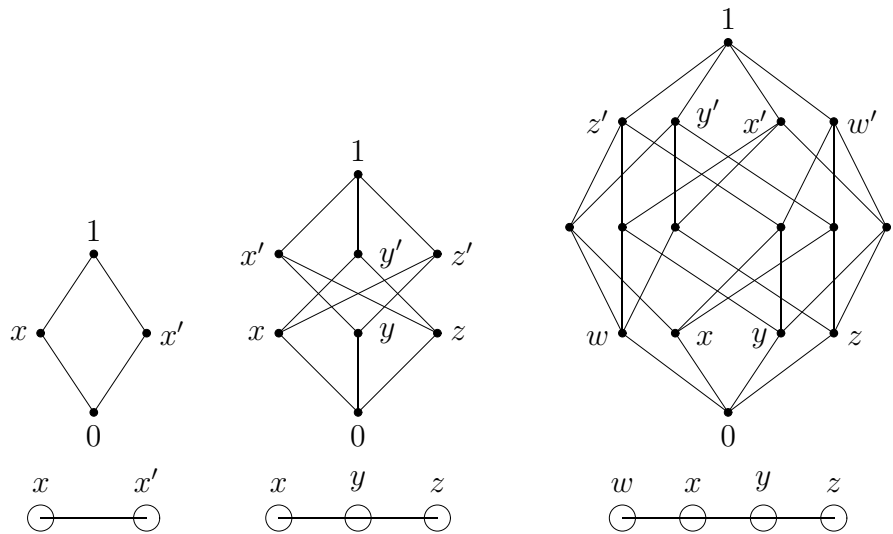
\begin{figure}[htbp]\centering
  % Set unitlength to default in case it's not
  \setlength{\unitlength}{1pt}
  \begin{picture}(330,190)(0,0)

    \put(0,0){
      % 2^2 Greechie diagram
      \begin{picture}(40,10)(0,0)
        \put(0,0){\line(1,0){40}}
        \put(0,0){\circle{10}}
        \put(40,0){\circle{10}}
        \put(0,10){\makebox(0,0)[b]{$x$}}
        \put(40,10){\makebox(0,0)[b]{$x'$}}
      \end{picture}
    }

    \put(0,40){
      % 2^2 Hasse diagram
      \begin{picture}(40,60)(0,0)
        \put(20,0){\line(-2,3){20}}
        \put(20,0){\line(2,3){20}}
        \put(20,60){\line(-2,-3){20}}
        \put(20,60){\line(2,-3){20}}

        \put(20,-5){\makebox(0,0)[t]{$0$}}
        \put(-5,30){\makebox(0,0)[r]{$x$}}
        \put(45,30){\makebox(0,0)[l]{$x'$}}
        \put(20,65){\makebox(0,0)[b]{$1$}}

        \put(20,0){\circle*{3}}
        \put(0,30){\circle*{3}}
        \put(40,30){\circle*{3}}
        \put(20,60){\circle*{3}}
      \end{picture}
    } % end of 2^2 subpicture

    \put(90,0){
      % 2^3 Greechie diagram
      \begin{picture}(60,10)(0,0)
        \put(0,0){\line(1,0){60}}
        \put(0,0){\circle{10}}
        \put(30,0){\circle{10}}
        \put(60,0){\circle{10}}
        \put(0,10){\makebox(0,0)[b]{$x$}}
        \put(30,10){\makebox(0,0)[b]{$y$}}
        \put(60,10){\makebox(0,0)[b]{$z$}}
      \end{picture}
    }

    \put(90,40){
      % 2^3 Hasse diagram
      \begin{picture}(60,90)(0,0)

        \put(30,0){\line(-1,1){30}}
        \put(30,0){\line(0,1){30}}
        \put(30,0){\line(1,1){30}}
        \put(30,90){\line(-1,-1){30}}
        \put(30,90){\line(0,-1){30}}
        \put(30,90){\line(1,-1){30}}
        \put(0,30){\line(1,1){30}}
        \put(30,30){\line(1,1){30}}
        \put(0,30){\line(2,1){60}}
        \put(30,30){\line(-1,1){30}}
        \put(60,30){\line(-1,1){30}}
        \put(60,30){\line(-2,1){60}}

        \put(30,-5){\makebox(0,0)[t]{$0$}}
        \put(-5,30){\makebox(0,0)[r]{$x$}}
        \put(38,30){\makebox(0,0)[l]{$y$}}
        \put(65,30){\makebox(0,0)[l]{$z$}}
        \put(-5,60){\makebox(0,0)[r]{$x'$}}
        \put(38,60){\makebox(0,0)[l]{$y'$}}
        \put(65,60){\makebox(0,0)[l]{$z'$}}
        \put(30,95){\makebox(0,0)[b]{$1$}}

        \put(0,30){\circle*{3}}
        \put(30,30){\circle*{3}}
        \put(60,30){\circle*{3}}
        \put(0,60){\circle*{3}}
        \put(30,60){\circle*{3}}
        \put(60,60){\circle*{3}}
        \put(30,0){\circle*{3}}
        \put(30,90){\circle*{3}}
      \end{picture}
    } % end of 2^3 subpicture

    \put(200,0){
      % 2^4 Greechie diagram
      \begin{picture}(100,10)(0,0)
        \put(20,0){\line(1,0){80}}
        \put(20,0){\circle{10}}
        \put(46.67,0){\circle{10}}
        \put(73.33,0){\circle{10}}
        \put(100,0){\circle{10}}
        \put(20,10){\makebox(0,0)[b]{$w$}}
        \put(46.67,10){\makebox(0,0)[b]{$x$}}
        \put(73.33,10){\makebox(0,0)[b]{$y$}}
        \put(100,10){\makebox(0,0)[b]{$z$}}
      \end{picture}
    }

    \put(200,40){
      % 2^4 Hasse diagram
      \begin{picture}(100,140)(0,0)
        \put(60,0){\line(-2,3){20}}
        \put(60,0){\line(2,3){20}}
        \put(60,0){\line(-4,3){40}}
        \put(60,0){\line(4,3){40}}
        \put(60,140){\line(-2,-3){20}}
        \put(60,140){\line(2,-3){20}}
        \put(60,140){\line(-4,-3){40}}
        \put(60,140){\line(4,-3){40}}

        \put(20,30){\line(-1,2){20}}
        \put(20,30){\line(0,1){40}}
        \put(20,30){\line(1,2){20}}
        \put(40,30){\line(-1,1){40}}
        \put(40,30){\line(1,1){40}}
        \put(40,30){\line(3,2){60}}
        \put(80,30){\line(-3,2){60}}
        \put(80,30){\line(0,1){40}}
        \put(80,30){\line(1,1){40}}
        \put(100,30){\line(-3,2){60}}
        \put(100,30){\line(0,1){40}}
        \put(100,30){\line(1,2){20}}

        \put(100,110){\line(1,-2){20}}
        \put(100,110){\line(0,-1){40}}
        \put(100,110){\line(-1,-2){20}}
        \put(80,110){\line(1,-1){40}}
        \put(80,110){\line(-1,-1){40}}
        \put(80,110){\line(-3,-2){60}}
        \put(40,110){\line(3,-2){60}}
        \put(40,110){\line(0,-1){40}}
        \put(40,110){\line(-1,-1){40}}
        \put(20,110){\line(3,-2){60}}
        \put(20,110){\line(0,-1){40}}
        \put(20,110){\line(-1,-2){20}}

        \put(60,140){\circle*{3}}
        \put(20,110){\circle*{3}}
        \put(40,110){\circle*{3}}
        \put(80,110){\circle*{3}}
        \put(100,110){\circle*{3}}
        \put(0,70){\circle*{3}}
        \put(20,70){\circle*{3}}
        \put(40,70){\circle*{3}}
        \put(80,70){\circle*{3}}
        \put(100,70){\circle*{3}}
        \put(120,70){\circle*{3}}
        \put(20,30){\circle*{3}}
        \put(40,30){\circle*{3}}
        \put(80,30){\circle*{3}}
        \put(100,30){\circle*{3}}
        \put(60,0){\circle*{3}}

        \put(60,-5){\makebox(0,0)[t]{$0$}}
        \put(105,28){\makebox(0,0)[l]{$z$}}
        \put(75,28){\makebox(0,0)[r]{$y$}}
        \put(48,28){\makebox(0,0)[l]{$x$}}
        \put(15,28){\makebox(0,0)[r]{$w$}}
        \put(105,110){\makebox(0,0)[l]{$w'$}}
        \put(75,110){\makebox(0,0)[r]{$x'$}}
        \put(48,113){\makebox(0,0)[l]{$y'$}}
        \put(15,110){\makebox(0,0)[r]{$z'$}}
        \put(60,145){\makebox(0,0)[b]{$1$}}

        %\put(-5,70){\makebox(0,0)[r]{$t$}}
        %\put(18,70){\makebox(0,0)[r]{$u$}}
        %\put(48,70){\makebox(0,0)[l]{$v$}}
        %\put(75,69){\makebox(0,0)[r]{$v'$}}
        %\put(104,70){\makebox(0,0)[l]{$u'$}}
        %\put(125,70){\makebox(0,0)[l]{$t'$}}

      \end{picture}
    } % end of 2^4 subpicture

  \end{picture}
  \caption{Greechie diagrams for Boolean lattices $2^2$, $2^3$, and $2^4$,
   labeled with the atoms of their corresponding Hasse diagrams shown
   above them.  ($2^4$ was adapted from \cite[Fig.~18, p.~84]{beran}.)
\label{fig:greechie-1}}
\end{figure}

\begin{figure}[htbp]\centering
  % Set unitlength to default in case it's not
  \setlength{\unitlength}{1pt}
  \begin{picture}(330,100)(0,0)

    \put(0,10){
% 123,345.
      \begin{picture}(60,80)(0,0)
        \put(0,30){\line(1,-1){30}}
        \put(30,0){\line(1,1){30}}
        \put(0,30){\circle{10}}
        \put(15,15){\circle{10}}
        \put(30,0){\circle{10}}
        \put(45,15){\circle{10}}
        \put(60,30){\circle{10}}
        \put(-10,30){\makebox(0,0)[r]{$v$}}
        \put(5,15){\makebox(0,0)[r]{$w$}}
        \put(30,-10){\makebox(0,0)[t]{$x$}}
        \put(55,15){\makebox(0,0)[l]{$y$}}
        \put(70,30){\makebox(0,0)[l]{$z$}}
      \end{picture}
    } % end of subpicture

    \put(100,0){
      \begin{picture}(60,80)(0,0)
        \qbezier(0,40)(30,60)(60,40)
        \put(30,0){\line(0,1){50}}
        \put(30,0){\circle{10}}
        \put(30,25){\circle{10}}
        \put(30,50){\circle{10}}
        \put(0,40){\circle{10}}
        \put(60,40){\circle{10}}
        \put(0,50){\makebox(0,0)[b]{$v$}}
        \put(30,60){\makebox(0,0)[b]{$x$}}
        \put(60,50){\makebox(0,0)[b]{$w$}}
        \put(40,25){\makebox(0,0)[l]{$y$}}
        \put(40,0){\makebox(0,0)[l]{$z$}}
      \end{picture}
    } % end of subpicture

    \put(200,5){
      % 2^3+2^3 Hasse diagram
      \begin{picture}(60,90)(0,0)
        \put(60,90){\line(-2,-1){60}}
        \put(60,90){\line(-1,-1){30}}
        \put(60,90){\line(0,-1){30}}
        \put(60,90){\line(1,-1){30}}
        \put(60,90){\line(2,-1){60}}
        \put(60,0){\line(-2,1){60}}
        \put(60,0){\line(-1,1){30}}
        \put(60,0){\line(0,1){30}}
        \put(60,0){\line(1,1){30}}
        \put(60,0){\line(2,1){60}}
        \put(0,60){\line(1,-1){30}}
        \put(0,60){\line(2,-1){60}}
        \put(30,60){\line(-1,-1){30}}
        \put(30,60){\line(1,-1){30}}
        \put(60,60){\line(-2,-1){60}}
        \put(60,60){\line(-1,-1){30}}
        \put(60,60){\line(1,-1){30}}
        \put(60,60){\line(2,-1){60}}
        \put(90,60){\line(-1,-1){30}}
        \put(90,60){\line(1,-1){30}}
        \put(120,60){\line(-2,-1){60}}
        \put(120,60){\line(-1,-1){30}}

        \put(60,-5){\makebox(0,0)[t]{$0$}}
        \put(-5,30){\makebox(0,0)[r]{$v$}}
        \put(38,30){\makebox(0,0)[l]{$w$}}
        \put(65,28){\makebox(0,0)[l]{$x$}}
        \put(96,30){\makebox(0,0)[l]{$y$}}
        \put(125,30){\makebox(0,0)[l]{$z$}}
        \put(-5,60){\makebox(0,0)[r]{$v'$}}
        \put(38,60){\makebox(0,0)[l]{$w'$}}
        \put(65,65){\makebox(0,0)[l]{$x'$}}
        \put(96,62){\makebox(0,0)[l]{$y'$}}
        \put(125,60){\makebox(0,0)[l]{$z'$}}
        \put(60,95){\makebox(0,0)[b]{$1$}}

        \put(60,0){\circle*{3}}
        \put(0,30){\circle*{3}}
        \put(30,30){\circle*{3}}
        \put(60,30){\circle*{3}}
        \put(90,30){\circle*{3}}
        \put(120,30){\circle*{3}}
        \put(0,60){\circle*{3}}
        \put(30,60){\circle*{3}}
        \put(60,60){\circle*{3}}
        \put(90,60){\circle*{3}}
        \put(120,60){\circle*{3}}
        \put(60,90){\circle*{3}}
      \end{picture}
    } % end of 2^3+2^3 subpicture

  \end{picture}
  \caption{Two different ways of drawing the same Greechie diagram, and its
    corresponding Hasse diagram.
\label{fig:greechie-2}}
\end{figure}
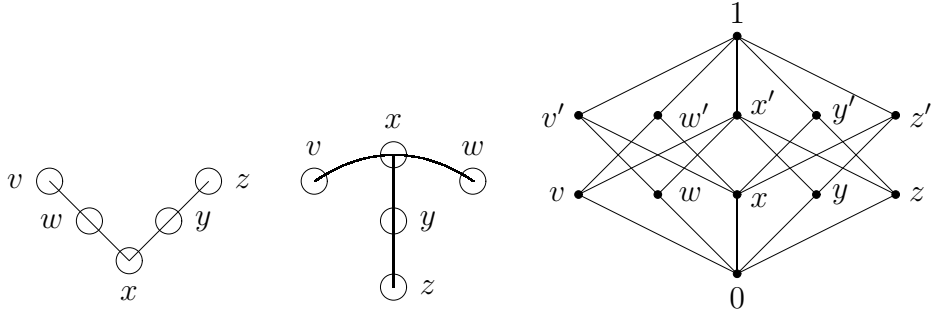

The Hasse diagrams for the Boolean algebras corresponding to 2-, 3-, and
4-atom blocks are shown in Fig.~\ref{fig:greechie-1}.
The Greechie diagram for a given lattice may be drawn in
several equivalent ways:  Fig.~\ref{fig:greechie-2}\ shows the same
Greechie diagram drawn in two different ways, along with the
corresponding Hasse diagram.  {}From the definitions we see that the
ordering of the atoms on a block does not matter, and we may also draw
blocks using arcs as well as straight lines as long as the blocks remain
clearly distinguishable.

Recall that a {\em poset\/} (partially ordered set) is a set with an
associated ordering relation that is reflexive ($a\le a$), antisymmetric
($a\le b,b\le a$ imply $a=b$), and transitive ($a\le b,b\le c$ imply
$a\le c$).  An {\em orthoposet\/} is a poset with lower and upper bounds
$0$ and $1$ and an operation~$'$ satisfying (i) if $a\le b$ then $b'\le
a'$; (ii) $a''=a$; and (iii) the infimum $a\cap a'$ and the supremum
$a\cup a'$ exist and are $0$ and $1$ respectively.  A {\em lattice\/} is
a poset in which any two elements have an infimum and a supremum.  An
orthoposet is {\em orthomodular\/} if $a\le b$ implies (i) the supremum
$a\cup b'$ exists and (ii) $a\cup(a'\cap b)=b$.  A lattice is {\em
orthomodular\/} if it is also an orthomodular poset.  For example,
Boolean algebras such as those of Fig.~\ref{fig:greechie-1} are
orthomodular lattices.  A {\em $\sigma$-orthomodular poset} is an
orthomodular poset in which every countable subset of elements has
a supremum.  An {\em atom\/} of an orthoposet is an element
$a\ne 0$ such that $b<a$ implies $b=0$.

In the literature, there are several different definitions of a Greechie
diagram.  For example, Beran (\cite[p.~144]{beran}) forbids 2-atom
blocks.  Kalmbach (\cite[p.~42]{kalmb83}) as well as Pt\'ak and
Pulmannov{\'a} \cite[p.~32]{ptak-pulm} include all diagrams with 2-atom
blocks connected to other blocks as long as the resulting pasting
corresponds to an orthoposet.  However, the case of 2-atom blocks
connected to other blocks is somewhat complicated; for example, the
definition of a loop in Def.~\ref{D:diagram} must be modified
(e.g.~\cite[p.~42]{kalmb83}) and no longer corresponds to the simple
geometry of a drawing of the diagram.  The definition of a Greechie
diagram also becomes more complicated; for example a pentagon (or any
$n$-gon with an odd number of sides) made out of 2-atom blocks is not a
Greechie diagram (i.e.\ does not correspond to any orthoposet).

The definition of Svozil and Tkadlec \cite{svozil-tkadlec} that we
adopt, Def.~\ref{D:greechie-diagram}, excludes 2-atom blocks connected
to other blocks.  It turns out that all orthomodular posets
representable by Kalmbach's definition can be represented with the
diagrams allowed by Svozil and Tkadlec's definition.  But the latter
definition eliminates the special treatment of 2-atom blocks connected
to other blocks and in particular simplifies any computer program
designed to process Greechie diagrams.

Svozil and Tkadlec's definition further restricts Greechie diagrams to
those diagrams representing orthoposets that are orthomodular by
forbidding loops of order less than~4, unlike the definitions of Beran
and Kalmbach.  The advantage appears to be mainly for convenience, as we
obtain only those Greechie diagrams that correspond to what are
sometimes called ``quantum logics'' ($\sigma$-orthomodular posets).
(We note that the term ``quantum logic'' is also used to denote a
propositional calculus based on orthomodular or weakly orthomodular
lattices. \cite{mpcommp99})

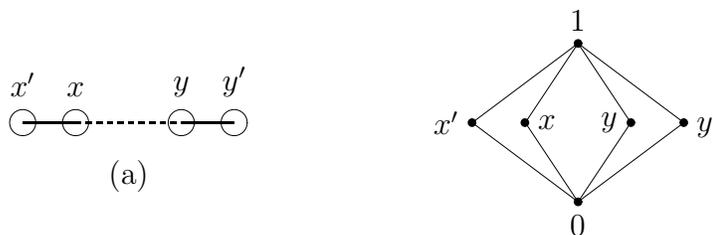
\begin{figure}[htbp]\centering
  % Set unitlength to default in case it's not
  \setlength{\unitlength}{1pt}
  \begin{picture}(240,80)(0,0)

    \put(0,30){
      % MO2 Greechie diagram
% 12,34.
      \begin{picture}(40,10)(0,0)
        \put(0,0){\line(1,0){20}}
        \put(0,0){\circle{10}}
        \put(20,0){\circle{10}}
        \put(0,10){\makebox(0,0)[b]{$x'$}}
        \put(20,10){\makebox(0,0)[b]{$x$}}
        \multiput(20,0)(4,0){10}{\line(1,0){2}}
        \put(60,0){\line(1,0){20}}
        \put(60,0){\circle{10}}
        \put(80,0){\circle{10}}
        \put(60,10){\makebox(0,0)[b]{$y$}}
        \put(80,10){\makebox(0,0)[b]{$y'$}}

        \put(40,-20){\makebox(0,0)[c]{(a)}}
      \end{picture}
    }

    \put(170,0){

      \begin{picture}(90,80)(0,0)
%        - 1 -
%      /  / \  \
%    x'  x   y  y'
%      \  \ /  /
%        - 0 -
        \put(40,0){\line(-2,3){20}}
        \put(40,0){\line(2,3){20}}
        \put(40,0){\line(-4,3){40}}
        \put(40,0){\line(4,3){40}}
        \put(40,60){\line(-2,-3){20}}
        \put(40,60){\line(2,-3){20}}
        \put(40,60){\line(-4,-3){40}}
        \put(40,60){\line(4,-3){40}}

        \put(40,-5){\makebox(0,0)[t]{$0$}}
        \put(25,30){\makebox(0,0)[l]{$x$}}
        \put(85,30){\makebox(0,0)[l]{$y'$}}
        \put(55,30){\makebox(0,0)[r]{$y$}}
        \put(-5,30){\makebox(0,0)[r]{$x'$}}
        \put(40,65){\makebox(0,0)[b]{$1$}}

        \put(40,0){\circle*{3}}
        \put(0,30){\circle*{3}}
        \put(20,30){\circle*{3}}
        \put(60,30){\circle*{3}}
        \put(80,30){\circle*{3}}
        \put(40,60){\circle*{3}}
      \end{picture}
    } % end of MO2 subpicture

  \end{picture}
  \caption{Greechie diagram for the lattice MO2 and its Hasse diagram.  The
dashed line indicates that the unconnected blocks belong to the same
Greechie diagram.
\label{fig:greechie-3}}
\end{figure}

The definition allows for Greechie diagrams whose blocks are not
connected.  In Fig.~\ref{fig:greechie-3} we show the Greechie diagram
for the {\em Chinese lantern\/} MO2 using unconnected 2-atom blocks.
This example also illustrates that even when the blocks are unconnected
the properties of the resulting orthoposet are not just a simple
combination of the properties of their components (as one might na\"\i
vely suppose), because we are adding disjoint sets of incomparable nodes
to the orthoposet.  As is well-known (\cite[p.~16]{kalmb83}), MO2 is not
distributive, unlike the Boolean blocks it is built from.

The {\tt latticeg} program takes, as its inputs, a Greechie diagram {\sc
ascii} representation (or more precisely a collection of them) and an
equation (or inference) to be tested.  This {\sc ascii} representation
is compatible with the output of the programs described in
Section~\ref{sec:generation}.  Currently {\tt latticeg} is designed to
work only with Greechie diagrams corresponding to lattices, i.e.\ that
have no loops of order~4 or less, as these are the most interesting for
studying equations valid in all Hilbert lattices.  It converts the
Greechie diagram to its corresponding Hasse diagram (internally stored as
truth tables).  Finally, the program tests all possible assignments of
nodes in the Hasse diagram to the input equation under test.

The {\tt latticeg} program also incorporates classical propositional
metalogic and predicate calculus to allow the study of such
characteristics as atomicity and superposition.

As a simple example of the operation of the {\tt latticeg} program, we
show how it verifies the passage and failure of the modular law [Equation
(\ref{eq:modular}) below] on the lattices of Figs.~\ref{fig:greechie-8}b and
\ref{fig:greechie-8}c.  We create a file with a name such
as {\tt m.gre} to represent
the lattices, containing the lines
\begin{verbatim}
       123,345.
       123,345,567.
\end{verbatim}
and run the program by typing
\begin{verbatim}
       latticeg -i m.gre "(av(b^(avc)))=((avb)^(avc))"
\end{verbatim}
The program responds with
\begin{verbatim}
       The input file has 2 lattices.
       Passed #1 (5/2/12)
       FAILED #2 (7/3/16) at (av(f^(avb)))=((avf)^(avb))
\end{verbatim}
The notation should be more or less apparent but is described in detail
by the program's help.  The numbers ``5/2/12'' show the atom/block/node
count, and the failure shows the internal Hasse diagram's nodal
assignment to the equation's variables.

Let us consider an application of the program. Closed subspaces
${\cal H}_a,{\cal H}_b$ of any infinite dimensional Hilbert
space ${\cal H}$ form a lattice in which the operations are
defined in the following way: $a'={\cal H}_a^\perp$,
$a\cap b\ =\ {\cal H}_a\bigcap {\cal H}_b$, and
$a\cup b\ =\ ({\cal H}_a+{\cal H}_b)^{\perp\perp}$.
In such a lattice its elements satisfy the following condition,
i.e., in any infinite dimensional Hilbert space its closed
subspaces satisfy the following equation which is called the
\it orthoarguesian equation\/\rm :
\begin{eqnarray}
\lefteqn{a \perp b \qquad \&\qquad c \perp d \qquad \&\qquad e
  \perp f \qquad \Rightarrow} & & \nonumber \\
& & ( a \cup b ) \cap ( c \cup d ) \cap ( e \cup f ) \le \nonumber \\
& &b \cup ( a \cap ( c \cup ( ( ( a \cup c )
\cap ( b \cup d ) ) \cap
( ( ( a \cup e ) \cap ( b \cup f ) ) \cup
( ( c \cup e ) \cap ( d \cup f ) ) ) ) ) )\quad \label{eq:6oa}
\end{eqnarray}
where $a\perp b$ means $a\le b'$.

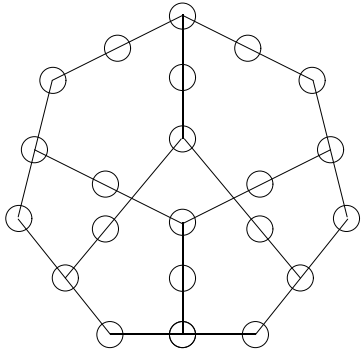
\begin{figure}[htbp]\centering
  \setlength{\unitlength}{1pt}
  \begin{picture}(360,150)(30,-10)
    \put(150,10){
        % heptagon sides (L46)
%DEM,BEI,AEJ,9CL,8CK,7CF,6DH,5DG,49B,38A,246,135,127.
        \put(110.9,96.2){\line(-2,1){49.3}}
        \put(110.9,96.2){\line(1,-4){13}}
        \put(89,0){\line(4,5){35}}
        \put(89,0){\line(-1,0){54.8}}
        \put(34.2,0){\line(-4,5){35}}
        \put(12.2,96.2){\line(-1,-4){13}}
        \put(12.2,96.2){\line(2,1){49.3}}
        % internal lines
        \put(61.5,74){\line(0,1){47}}
        \put(17.1,21.4){\line(5,6){44}}
        \put(106.0,21.4){\line(-5,6){44}}
        \put(5.5,70){\line(2,-1){56}}
        \put(117.5,70){\line(-2,-1){56}}
        \put(61.5,0){\line(0,1){42.5}}
        % heptagon vertices (L46)
        \put(61.5,120.5){\circle{10}}
        \put(110.5,96.2){\circle{10}}
        \put(123.5,44){\circle{10}}
        \put(89,0){\circle{10}}
        \put(34,0){\circle{10}}
        \put(-0.5,44){\circle{10}}
        \put(12.5,96.2){\circle{10}}
        % heptagon mid-sides
        \put(86.2,108.4){\circle{10}}
        \put(117.5,70){\circle{10}}
        \put(106.0,21.4){\circle{10}}
        \put(61.5,0){\circle{10}}
        \put(17.1,21.4){\circle{10}}
        \put(5.5,70){\circle{10}}
        \put(36.9,108.4){\circle{10}}
        % internal nodes
        \put(32.3,57){\circle{10}}
        \put(90.7,57){\circle{10}}
        \put(61.5,0){\circle{10}}
        \put(61.5,21.25){\circle{10}}
        \put(61.5,42.5){\circle{10}}
        \put(61.5,97.3){\circle{10}}
        \put(61.5,74){\circle{10}}
        \put(32.3,40){\circle{10}}
        \put(90.7,40){\circle{10}}}
      \end{picture}
  \caption{\hbox to3mm{\hfill}
   Greechie diagram for OML L46.
\label{fig:5oa}}
\end{figure}

\begin{figure}[htbp]\centering
  % Set unitlength to default in case it's not
  \setlength{\unitlength}{1pt}
  \begin{picture}(240,140)(0,0)

    \put(5,23) { % argument of \put is offset of Peterson in figure
      \begin{picture}(124,110)(0,0) % start Peterson subpicture
% 123,345,567,789,9AB,BC1,2E8,4FA,6DC,DEF.
        % hexagon sides (Peterson)
        \put(32.2,0){\line(1,0){55.6}}
        \put(32.2,100){\line(1,0){55.6}}
        \put(2.3,50){\line(3,5){29.9}}
        \put(117.7,50){\line(-3,5){29.9}}
        \put(2.3,50){\line(3,-5){29.9}}
        \put(117.7,50){\line(-3,-5){29.9}}
        % internal lines
        \put(60,100){\line(0,-1){100}}
        \put(17.25,25){\line(5,3){84.8}}
        \put(102.75,25){\line(-5,3){84.8}}
        \put(34.71,65.6){\line(1,0){50.58}}

        % hexagon atoms (Peterson)
        \put(2.3,50){\circle{10}}
        \put(117.7,50){\circle{10}}
        \put(87.8,0){\circle{10}}
        \put(87.8,100){\circle{10}}
        \put(32.2,0){\circle{10}}
        \put(32.2,100){\circle{10}}
        % hexagon mid-sides
        \put(60,0){\circle{10}}
        \put(60,100){\circle{10}}
        \put(17.25,25){\circle{10}}
        %\put(17.25,75){\circle{10}} original symmetrical
        \put(17.85,76){\circle{10}}
        \put(102.75,25){\circle{10}}
        %\put(102.75,75){\circle{10}} original symmetrical
        \put(102.15,76){\circle{10}}
        % internal nodes
        \put(34.71,65.6){\circle{10}}
        \put(85.29,65.6){\circle{10}}
        \put(60,65.6){\circle{10}}
        \put(60,-20){\makebox(0,0)[c]{(a)}}
      \end{picture}
    } % end of Peterson subpicture

    \put(170,90) {
      \begin{picture}(50,60)(0,0)
        \put(0,30){\line(1,-1){30}}
        \put(30,0){\line(1,0){30}}
        \put(0,30){\circle{10}}
        \put(15,15){\circle{10}}
        \put(30,0){\circle{10}}
        \put(45,0){\circle{10}}
        \put(60,0){\circle{10}}
        \put(45,-20){\makebox(0,0)[c]{(b)}}
      \end{picture}
    }

    \put(170,30) {
      \begin{picture}(50,60)(0,0)
        \put(0,30){\line(1,-1){30}}
        \put(30,0){\line(1,0){30}}
        \put(60,0){\line(1,1){30}}
        \put(0,30){\circle{10}}
        \put(15,15){\circle{10}}
        \put(30,0){\circle{10}}
        \put(45,0){\circle{10}}
        \put(60,0){\circle{10}}
        \put(75,15){\circle{10}}
        \put(90,30){\circle{10}}
        \put(45,-20){\makebox(0,0)[c]{(c)}}
      \end{picture}
    }

  \end{picture}
  \caption{(a)
    Peterson OML. (b) Greechie diagram obeying modular law.
    (c) Greechie diagram violating modular law.
\label{fig:greechie-8}}
\end{figure}
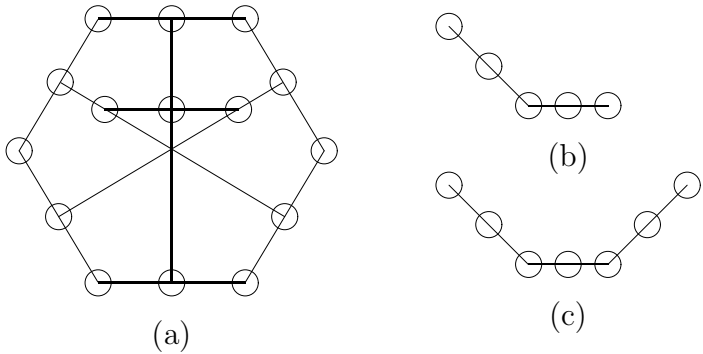

We wanted first, to reduce the number of variables in this equation and
second, to generalize the equation to $n$ variables.  In attacking the
first problem the program helped us to quickly eliminate dead ends:  a
failure of a conjectured equation in a lattice in which Equation
\ref{eq:6oa} held meant that the latter equation was weaker and
vice-versa.  Thus we arrived at the following 4-variable
equation---which we call the 4OA law:
\begin{eqnarray}
(a_1\to_1 a_3) \cap (a_1{\buildrel (4)\over\equiv}a_2)
\le a_2\to_1 a_3\,.\label{eq:4oa}
\end{eqnarray}
where the operation ${\buildrel (4)\over\equiv}$ is defined as
follows:
\begin{eqnarray}
a_1{\buildrel (4)\over\equiv}a_2\
&{\buildrel\rm def\over =}&\
\ (a_1{\buildrel (3)\over\equiv}a_2)\cup
((a_1{\buildrel (3)\over\equiv}a_4)\cap
(a_2{\buildrel (3)\over\equiv}a_4))\,,\label{def:4oa}
\end{eqnarray}
where
\begin{eqnarray}
a_1{\buildrel (3)\over\equiv}a_2\
&{\buildrel\rm def\over =}&\
((a_1\to_1 a_3)\cap(a_2\to_1 a_3))
\cup((a_1'\to_1 a_3)\cap(a_2'\to_1 a_3)),
\end{eqnarray}
where $a\to_1b\ {\buildrel\rm def\over =}\ a'\cup(a\cap b)$.
We then proved ``by hand'' that Equations (\ref{eq:6oa}) and
(\ref{eq:4oa}) are equivalent. \cite{mpoa99} We also proved that the
following generalization (which we call the $n$OA law)
of Equation (\ref{eq:4oa})
\begin{eqnarray}
(a_1\to_1 a_3) \cap (a_1{\buildrel (n)\over\equiv}a_2)
\le a_2\to_1 a_3\label{eq:noa}
\end{eqnarray}
where
\begin{eqnarray}
a_1{\buildrel (n)\over\equiv}a_2\quad{\buildrel\rm def\over =}\quad
(a_1{\buildrel (n-1)\over\equiv}a_2)\cup
((a_1{\buildrel (n-1)\over\equiv}a_n)\cap
(a_2{\buildrel (n-1)\over\equiv}a_n))\,,\quad n\ge 4
\end{eqnarray}
holds in any Hilbert lattice.  To show that this generalization is
a nontrivial one, the program is all we need---we need not
prove anything ``by hand.'' It suffices to find a
Greechie diagram in which the 4OA law holds and 5OA law fails.
An 800 MHz PC took a few days to find such a lattice (shown in
Fig.~\ref{fig:5oa}), \cite{mpoa99} while to find it ``by hand'' is, due
to the number of variables in the equation and nodes in the
corresponding Hasse diagram, apparently humanly impossible.

Considerable effort was put into making the program run fast, with
methods such as exiting an evaluation early when a hypothesis of an
inference fails.  Truth tables for all built-in compound operations
(such as the various quantum implications and the quantum biconditional)
are precomputed.  The innermost loop (which evaluates an assignment) was
optimized for the fastest runtime we could achieve.  The number of
assignments of lattice nodes to equation variables that must be tested
is $n^v$ where $n$ is the number of nodes in the Hasse diagram and $v$
is the number of variables in the equation.  The algorithm requires a
time approximately proportional to $kn^v$ where $k$ is the length of the
equation expressed in Polish notation.  For a typical equation ($k =
20$) with no hypotheses, the algorithm currently evaluates around a
million assignments per second on an 800-MHz PC.  The speed is typically
faster when hypotheses are present, particularly if they have fewer
variables than the conclusion.  If a lattice violates an equation, it
often happens (with luck) that the first failure will be found quickly,
in which case further evaluations do not have to be done.

The propositional metalogic feature of {\tt latticeg}, when
carefully used in a series of hypotheses with a successively
increasing number of variables, can sometimes be exploited to achieve
orders of magnitude speed-up with certain equations containing many
variables.  For example, we tested the 8-variable Godowski equation
against a 42-node lattice in 16 hours, whereas without the speed-up it
would have required around $10^5$ hours.  To illustrate how this
speed-up works, we can add to the 4-variable Godowski equation
[Equation~\ref{eq:4go} below] hypotheses as follows:
\begin{eqnarray}
\lefteqn{\sim (d\to_1a\le a\to_1d) \qquad \& \qquad
 \sim ((c\to_1d)\cap(d\to_1a)\le a\to_1d)
 \qquad \Rightarrow} & & \nonumber \\
& &
\qquad\qquad\qquad\qquad\qquad
 (a\to_1b)\cap(b\to_1c)\cap(c\to_1d)\cap(d\to_1a)
\le a\to_1d\,.\label{eq:4gohyp}
\end{eqnarray}
Here $\sim$ means metalogical {\sc not}.  The hypotheses are redundant
in any ortholattice, which is the case for the Greechie diagrams that
are of interest to us.  We take advantage of the {\tt latticeg} feature
that exits the evaluation of a lattice nodal assignment if a hypothesis
fails.  If an assignment to the first hypothesis, with only 2 variables,
fails (as it typically does for some assignments) it means we don't have
to scan the remaining variables.  Similarly, if the first hypothesis
passes but the second (with 3 variables) fails, we can skip the
evaluation of the 4-variable conclusion.

We are currently exploring the exploitation of possible symmetries
inherent in the Gree\-chie diagram to speed up the program further, but
we have not yet achieved any results in this direction.  For certain
special cases such as the Godowski equations we are also exploring the
use of a ``dynamic programming'' technique that may provide a run
time proportional to $kn^4$ instead of $kn^v$, regardless of the number
of variables.

For fastest run time, it is desirable to screen equations with the
smallest Greechie diagrams first.  For this purpose what matters is the
size of the Hasse diagram and not the Greechie diagram.  In a chain of
blocks each having two atoms connected, a 3-atom block adds 4 nodes to
the Hasse diagram whereas a 4-atom block adds 12 nodes.  For example,
the decagon (10 blocks) has 42 nodes with 3-atom blocks and 122 nodes
with 4-atom blocks.  With 3-atom blocks, a 6-variable equation---our
practical upper limit when there are no strong hypotheses---must be
evaluated $42^6 = 5.5$ billion times (a few hours of CPU time on an
800-MHz PC), but with 4-atom blocks it would take $122^6 = 3.3$ trillion
evaluations, which is currently impractical.  So far most of our work
has been done using diagrams with every block having size 3.

Two other heuristics have helped us to falsify
conjectures more quickly.  The first is to first scan Greechie diagrams
with the highest block-to-atom ratio (Table~\ref{T:tableA}).  Such
diagrams seem to have the most complex ``structure'' with the most
likelihood of violating a non-orthomodular equation (``non-orthomodular
equation'' here means an equation which turns the orthomodular lattice
variety into a smaller one when added to it).  A drawback is that at
higher atom counts, virtually every such diagram violates almost any
non-orthomodular equation, making it very useful for identifying
non-orthomodular properties but less useful for proving independence
results.
For example, for 35 and 36 atoms (the case we elaborated in
Section~\ref{sec:generation}) the highest ratio is 1
(Table~\ref{T:tableB}) and in those diagrams all equations that we
know to be non-orthomodular fail. Hence, for example $36\times36$
(36 atoms, 36 blocks) is a very useful tool for an initial scanning
of equations we want to check for ``non-orthomodularity.''

The second heuristic is our empirical observation that Greechie diagrams
without feet often behave the same as the same diagram with feet
added.  For example, the Peterson OML (Fig.~\ref{fig:greechie-8}a), with
32 nodes, is the smallest lattice that violates Godowski's 4-variable
strong state law \cite{mpoa99} which holds in any infinite dimensional
Hilbert space:
\begin{eqnarray}
(a\to_1b)\cap(b\to_1c)\cap(c\to_1d)\cap(d\to_1a)
&\le&a\to_1d\label{eq:4go}
\end{eqnarray}
but not Godowski's 3-variable law (which also holds in infinite
dimensional Hilbert spaces)
\begin{eqnarray}
(a\to_1b)\cap(b\to_1c)\cap(c\to_1a)
&\le&a\to_1c\,.
\label{eq:3go}
\end{eqnarray}
The Peterson OML is useful as a test for an equation derived from the
4-variable law and conjectured to be equivalent to it:  if it does not
violate the conjectured equation we know the equation is weaker the 4-variable
law.  Now, we observe empirically that we may add a foot (a 3-atom block
connected at only one point) to any of its 15 atoms without changing
this behavior.  We have also not seen a chain of feet or combination of
feet that changes this behavior when added to the diagram.

So, by scanning only lattices without feet, we can obtain a speedup of
20 times for 14-block lattices (Table~\ref{T:tableA}).  Supporting this
heuristic is the fact that complex Greechie diagrams with feet are rarely
found in the literature. We obtained an additional support by scanning
the 4OA law given by Equation (\ref{eq:4oa}) and Godowski's 3
variable equation (\ref{eq:3go}) through several million lattices
with free feet, vs.\ those with feet stripped: we did not find a single
difference. To our knowledge there is only one  special case 
for which feet do make a difference. Fig.~\ref{fig:greechie-8}b shows 
a lattice that obeys the modular law
\begin{eqnarray}
a\cup(b\cap(a\cup c))
&=&(a\cup b)\cap(a\cup c)
\label{eq:modular}
\end{eqnarray}
but violates it when a foot is added (Fig.~\ref{fig:greechie-8}c).
This special case might well be insignificant because of all Greechie 
diagrams only star-like ones (Fig.~\ref{fig:greechie-8}b, $D_{3,2}$ and 
$D_{4,4}$ from Fig.~\ref{fig:tree}, etc.) are modular. As soon 
as we add any block to any other atom apart from the central one in  
such a lattice we make it non-modular. 

Another heuristic for reducing the number of diagrams to be scanned is
suggested by the following observation, although we have not implemented
it.  We can list the diagrams in such a way that (except for the first
diagram) each is formed by adding one block to a diagram earlier in the
list.  Whenever the earlier diagram violates an equation, we have
observed that it is very likely that the new diagram will also violate
the equation.  By skipping the new diagram in this case (presuming the
probable violation), a speedup can be obtained.

\begin{figure}[htbp]\centering
  % Set unitlength to default in case it's not
  \setlength{\unitlength}{1pt}
  \begin{picture}(260,150)(-10,-10)

    \put(0,0) { % argument of \put is offset of L38 in figure
      \begin{picture}(124,120)(0,0) % start L38 subpicture
% 123,345,567,789,9AB,BCD,DE1,CF4,FGH,HI6.
% alternate version:
% 9CF,8BG,7AD,6AE,5BH,4CI,279,235,168,134.
        % heptagon sides (L38)
        \put(110.9,96.2){\line(-2,1){49.3}}
        \put(110.9,96.2){\line(1,-4){13}}
        \put(89,0){\line(4,5){35}}
        \put(89,0){\line(-1,0){54.8}}
        \put(34.2,0){\line(-4,5){35}}
        \put(12.2,96.2){\line(-1,-4){13}}
        \put(12.2,96.2){\line(2,1){49.3}}
        % internal lines
        \put(5.5,70){\line(1,0){112}}
        \put(61.5,70){\line(0,-1){41.2}}
        \put(61.5,28.8){\line(6,-1){44.5}}

        % heptagon vertices (L38)
        \put(61.5,120.5){\circle{10}}
        \put(110.5,96.2){\circle{10}}
        \put(123.5,44){\circle{10}}
        \put(89,0){\circle{10}}
        \put(34,0){\circle{10}}
        \put(-0.5,44){\circle{10}}
        \put(12.5,96.2){\circle{10}}
        % heptagon mid-sides
        \put(86.2,108.4){\circle{10}}
        \put(117.5,70){\circle{10}}
        \put(106.0,21.4){\circle{10}}
        \put(61.5,0){\circle{10}}
        \put(17.1,21.4){\circle{10}}
        \put(5.5,70){\circle{10}}
        \put(36.9,108.4){\circle{10}}
        % internal nodes
        \put(61.5,70){\circle{10}}
        \put(61.5,49.2){\circle{10}}
        \put(61.5,28.8){\circle{10}}
        \put(83.8,25.1){\circle{10}}
      \end{picture}
    } % end of L38 subpicture

    \put(200,0) { % argument of \put is offset of L38+2 in figure
      \begin{picture}(124,120)(0,0) % start L38+2 subpicture
% 123,345,567,789,9AB,BCD,DE1,CF4,FGH,HI6,AHJ,1K8.
        % heptagon sides (L38)
        \put(110.9,96.2){\line(-2,1){49.3}}
        \put(110.9,96.2){\line(1,-4){13}}
        \put(89,0){\line(4,5){35}}
        \put(89,0){\line(-1,0){54.8}}
        \put(34.2,0){\line(-4,5){35}}
        \put(12.2,96.2){\line(-1,-4){13}}
        \put(12.2,96.2){\line(2,1){49.3}}
        % internal lines
        \put(5.5,70){\line(1,0){112}}
        \put(61.5,70){\line(0,-1){41.2}}
        \put(61.5,28.8){\line(6,-1){44.5}}
        \put(61.5,28.8){\line(-6,-1){44.5}}
        \qbezier(61.5,0)(93,60.25)(61.5,120.5)

        % heptagon vertices (L38)
        \put(61.5,120.5){\circle{10}}
        \put(110.5,96.2){\circle{10}}
        \put(123.5,44){\circle{10}}
        \put(89,0){\circle{10}}
        \put(34,0){\circle{10}}
        \put(-0.5,44){\circle{10}}
        \put(12.5,96.2){\circle{10}}
        % heptagon mid-sides
        \put(86.2,108.4){\circle{10}}
        \put(117.5,70){\circle{10}}
        \put(106.0,21.4){\circle{10}}
        \put(61.5,0){\circle{10}}
        \put(17.1,21.4){\circle{10}}
        \put(5.5,70){\circle{10}}
        \put(36.9,108.4){\circle{10}}
        % internal nodes
        \put(61.5,70){\circle{10}}
        \put(61.5,49.2){\circle{10}}
        \put(61.5,28.8){\circle{10}}
        \put(83.8,25.1){\circle{10}}
        \put(39.2,25.1){\circle{10}}
        \put(77.5,58){\circle{10}}
      \end{picture}
    } % end of L38+2 subpicture

  \end{picture}
  \caption{\hbox to3mm{\hfill}(a)
   Greechie diagram for L38;
  \hbox to5mm{\hfill} (b) L38 with two blocks added.
\label{fig-oaaddegdes}}
\end{figure}
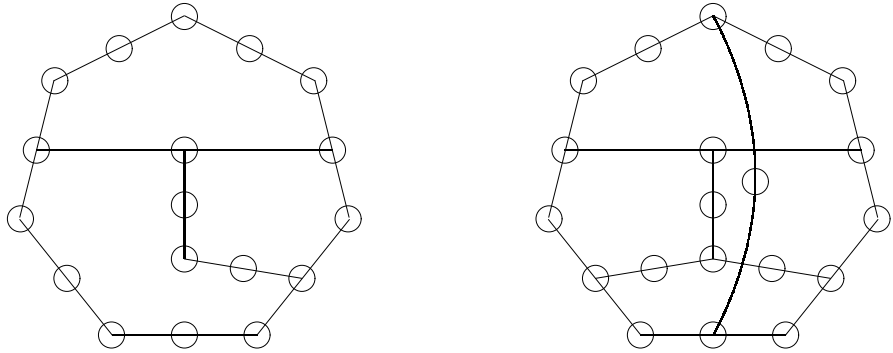

However the above observation does not universally hold, i.e.\ sometimes
the new diagram will pass an equation violated by the earlier one.  An
example is shown in Fig.~\ref{fig-oaaddegdes}.  The diagram L38 violates
the orthoarguesian law (Equation~\ref{eq:6oa}).  But if we extend L38 by
adding two blocks as shown in Fig.~\ref{fig-oaaddegdes}b, the resulting
diagram will pass not only this law [equivalent to 4OA given by 
Eq.~(\ref{eq:4oa})] but also 5OA and 6OA [given by 
Eq.~(\ref{eq:noa}) for $n=5$ and $n=6$, respectively].  

If our various speedup heuristics are used for practical reasons, the
user must be aware that a diagram scan may be incomplete.  So if, in
our example, the extended L38 (by passing) could serve to prove a
certain independence result, it would be missed by the scan.  Currently
we have no data to indicate how often such cases would be missed.

Any scan of diagrams to test an equation is, of course, {\em a priori}
incomplete since the number of diagrams is infinite.  The various
heuristics we have described may cause some lattices in any finite list
to be skipped.  But if a lattice with the desired properties is found
more quickly our goal is achieved.  Of course a scan can be continued
for any diagrams omitted by the heuristics if more completeness is
desired.

\section{Conclusions}
\label{sec:concl}

Greechie diagrams (we used the version given by Definition 
\ref{D:diagram} and discussed the others in Section \ref{sec:latticeg}) 
are generators of examples and counterexamples of orthomodular
non-modular lattices. They are special cases of lattices one 
obtains by using the recent generalization by Navara and Rogalewicz 
\cite{rogal,navara-rogal} of Dichtl's pasting construction for
orthomodular posets and lattices \cite{dichtl}.  
Navara and Rogalewicz's method exhaustively generates all finite 
orthomodular non-modular lattices but Greechie diagrams are 
apparently easier to generate and certainly much easier to test 
lattice equations. For these reasons, Greechie diagrams have been used 
almost exclusively so far.    

Since any infinite dimensional complex Hilbert space is orthoisomorphic 
to a Hilbert lattice which is orthomodular and non-modular, Greechie 
diagrams represent an indispensable tool for Hilbert space 
investigation. This has also been prompted by recent developments in 
the field of quantum computing. However, as we stressed in the 
Introduction, the existing (both manual and automated) constructions of
Greechie diagrams and their application to Hilbert space properties
(which resulted in many important results at the time) recently
reached the frontiers of human manageability. Therefore, in
Section \ref{sec:generation} we gave an algorithm and a program
for generating Greechie lattices with theoretically unlimited numbers
of atoms and blocks. The algorithm is of a completely
different kind from the only earlier algorithm and the program
is at least $10^5$ times faster.
In Section \ref{sec:latticeg} we then gave an algorithm and programs
for automated checking of lattice properties on Greechie diagrams.

Our algorithm for generating Greechie diagrams---given by Definition
\ref{def:procedure}---works by defining a unique construction path
for each isomorphism class of diagrams. It enabled us
to produce a self-contained program {\tt greechie} for automated
generation of diagrams with a specified number or range of atoms and/or
blocks. Several properties and several special types of Greechie
diagram construction are discussed in the Section, as are
connections to some equivalent results in graph theory.

The algorithm for automated checking of lattice properties described in
Section \ref{sec:latticeg} works by converting a Greechie diagram to
its corresponding Hasse diagram and then converting the Hasse diagram to a
truth table for supremum and orthocomplementation.  It enabled us to
construct a self-contained program {\tt latticeg} which takes, as its
inputs, Greechie diagrams in {\sc ascii} representation and an equation
or inference or quantified expression to be tested.  Many
programming speed-ups have been used to make the program run as fast as
possible.  For example, all built-in compound operations are
precomputed, many C code tricks are used, etc.  Also, to additionally
speed up scanning, we use many Greechie diagram heuristics that we found
recently:  a lattice equation is most likely to fail in a lattice with
the highest block-to-atom ratio, scanning of an equation on a diagram
with no feet and on the same diagram with feet added makes no difference
for most diagrams, etc.  Several properties and several
special cases of Greechie lattices are given and discussed in the
Section.  In particular, it is explained how we proved that our recent
$n$-variable generalization of the orthoarguesian equation is a
non-trivial one by using nothing but {\tt latticeg} applied to an output
of {\tt greechie}.

\bigskip
\bibliographystyle{report}

\begin{thebibliography}{10}

\bibitem{birk-v-neum}
G.~Birkhoff and J.~{{v}on Neumann},
\newblock The Logic of Quantum Mechanics,
\newblock Ann. Math. {\bf {\bf 37}}, 823--843 (1936).

\bibitem{mackey}
G.~W. Mackey,
\newblock {\em The Mathematical Foundations of Quantum Mechanics},
\newblock W. A. Benjamin, New York, 1963.

\bibitem{zierler}
N.~Zierler,
\newblock Axioms for Non-Relativistic Quantum Mechanics,
\newblock Pacif. J. Math. {\bf {\bf 11}}, 1151--1169 (1961).

\bibitem{piron}
C.~Piron,
\newblock Axiomatique quantique,
\newblock Helv. Phys. Acta {\bf {\bf 37}}, 439--468 (1994).

\bibitem{maclaren}
M.~D. Mac{L}aren,
\newblock Atomic Orthocomplemented Lattices,
\newblock Pacif. J. Math. {\bf {\bf 14}}, 597--612 (1964).

\bibitem{araki}
I.~Amemiya and H.~Araki,
\newblock A Remark on {P}iron's Paper,
\newblock Publ. Research Inst. Math. Sci., Kyoto Univ. Series {\bf {\bf A
  2}}(3), 423--427 (1966/67).

\bibitem{maczin}
M.~J. M{\c a}czy{\'n}ski,
\newblock Hilbert Space Formalism of Quantum Mechanics without the {H}ilbert
  Space Axiom,
\newblock Rep. Math. Phys. {\bf {\bf 3}}, 209--219 (1972).

\bibitem{keller}
H.~A. Keller,
\newblock Ein {n}icht-{k}lassischer {H}ilbertscher {R}aum,
\newblock Math. Z. {\bf {\bf 172}}, 41--49 (1980).

\bibitem{soler}
M.~P. Sol{\`e}r,
\newblock Characterization of {H}ilbert Spaces by Orthomodular Spaces,
\newblock Comm. Alg. {\bf {\bf 23}}, 219--243 (1995).

\bibitem{holl95}
S.~S. {Holland, JR.},
\newblock Orthomodularity in Infinite Dimensions; a Theorem of {M}.
  {S}ol{\`e}r,
\newblock Bull. Am. Math. Soc. {\bf {\bf 32}}, 205--234 (1995).

\bibitem{prestel}
A.~Prestel,
\newblock On {S}ol\`er's Characterization of {H}ilbert Spaces,
\newblock Manus. Math. {\bf {\bf 86}}, 225--238 (1995).

\bibitem{mayet98}
R.~Mayet,
\newblock Some Characterizations of Underlying Division Ring of a {H}ilbert
  Lattice of Authomorphisms,
\newblock Int. J. Theor. Phys. {\bf {\bf 37}}, 109--114 (1998).

\bibitem{dvurecen96}
A.~Dvure{\v c}enskij,
\newblock Test Spaces and Characterizations of Quadratic Spaces,
\newblock Int. J. Theor. Phys. {\bf {\bf 35}}, 2093--2106 (1996).

\bibitem{dvurecen98}
A.~Dvure{\v c}enskij,
\newblock Sol{\`e}r Theorem and Characterization of Inner Product Spaces,
\newblock Int. J. Theor. Phys. {\bf {\bf 37}}, 23--29 (1998).

\bibitem{mphpa98}
M.~Pavi{\v c}i{\'c} and N.~D. Megill,
\newblock Binary Orthologic with Modus Ponens Is either Orthomodular or
  Distributive,
\newblock Helv. Phys. Acta {\bf {\bf 71}}, 610--628 (1998).

\bibitem{greechie78}
R.~J. Greechie,
\newblock Another Nonstandard Quantum Logic (and how {I} Found It),
\newblock in {\em Mathematical Foundations of Quantum Theory (Papers from a
  conference held at Loyola University, New Orleans, June 2--4, 1977)}, edited
  by A.~R. Marlow, pages 71--85, Academic Press, New York, 1978.

\bibitem{kalmb83}
G.~Kalmbach,
\newblock {\em Orthomodular Lattices},
\newblock Academic Press, London, 1983.

\bibitem{mckay98}
B.~D. Mc{K}ay,
\newblock Isomorph-Free Exhaustive Generation,
\newblock J. Algorithms {\bf {\bf 26}}, 306--324 (1998).

\bibitem{mpoa99}
N.~D. Megill and M.~Pavi{\v c}i{\'c},
\newblock Equations and State Properties That Hold in All Closed Subspaces of
  an Infinite Dimensional {H}ilbert Space,
\newblock Int. J. Theor. Phys., {\bf 39}, No.~10 (2000).

\bibitem{svozil-tkadlec}
K.~Svozil and J.~Tkadlec,
\newblock Greechie Diagrams, Nonexistence of Measures and
  {K}ochen-{S}pecker-Type Constructions,
\newblock J. Math. Phys. {\bf 37}, 5380--5401 (1996).

\bibitem{ptak-pulm}
P.~Pt\'ak and S.~Pulmannov{\'a},
\newblock {\em Orthomodular Structures as Quantum Logics},
\newblock Kluwer, Dordrecht, 1991.

\bibitem{mckay90}
B.~D. Mc{K}ay,
\newblock {\tt {n}auty} User's Guide (version 1.5),
\newblock Dept. Computer Science, Australian Nat. Univ. {\bf Tech. Rpt.
  {TR}-{CS}-90-02} (1990).

\bibitem{OKeefe}
M.~O'Keefe and P.~K. Wong,
\newblock A Smallest Graph with Girth 10 and Valency 3,
\newblock J. Combinatorial Theory, Series B {\bf {\bf 29}}, 91--105 (1980).

\bibitem{mmn98}
B.~D. Mc{K}ay, W.~Myrvold, and J.~Nadon,
\newblock Fast backtracking principles applied to find new cages,
\newblock in {\em Ninth Annual {ACM-SIAM} Symposium on Discrete Algorithms},
  pages 188--191, SIAM, New York, 1998.

\bibitem{beran}
L.~Beran,
\newblock {\em Orthomodular Lattices; Algebraic Approach},
\newblock D.~Reidel, Dordrecht, 1985.

\bibitem{mpcommp99}
M.~Pavi{\v c}i{\'c} and N.~D. Megill,
\newblock Non-Orthomodular Models for Both Standard Quantum Logic and Standard
  Classical Logic: Repercussions for Quantum Computers,
\newblock Helv. Phys. Acta {\bf {\bf 72}}, 189--210 (1999),
\newblock http://xxx.lanl.gov/abs/quant-ph/9906101.

\bibitem{rogal}
V.~Rogalewicz,
\newblock Any orthomodular poset is a pasting of Boolean algebras,
\newblock Comment. Math. Univ. Carolin. {\bf {\bf 29}}, 557--558 (1988).

\bibitem{navara-rogal}
M.~Navara and V.~Rogalewicz,
\newblock The Pasting Constructions for Orthomodular Posets,
\newblock Math. Nachr. {\bf {\bf 154}}, 157--168 (1991).

\bibitem{dichtl}
M.~Dichtl,
\newblock Astroids and Pasting,
\newblock Algebra Universalis {\bf {\bf 18}}, 380--385 (1981).

\end{thebibliography}

\end{document}